\documentclass[showpacs,superscriptaddress,preprint,prb]{revtex4} 
\usepackage{amsmath,graphicx}
\usepackage{subfigure} 
\usepackage{braket}
\usepackage{subfigure}
\usepackage[usenames]{color}
\usepackage[dvipsnames]{xcolor}
\usepackage{ulem}
\usepackage{xcolor,soul}
\usepackage[T1]{fontenc}

\begin{document}
\title {Tunnelling spectroscopy of few-monolayer NbSe$_2$ in high magnetic field: triplet superconductivity and Ising protection}
\author{M. Kuzmanović}
\thanks{These two authors contributed equally.}
\affiliation{Laboratoire de Physique des Solides (CNRS UMR 8502), Bâtiment 510, Université Paris-Saclay 91405 Orsay, France}  
\affiliation{QTF Centre of Excellence, Department of Applied Physics, Aalto University School of Science, P.O. Box 15100, 00076 Aalto, Finland}
\author{T. Dvir} 
\thanks{These two authors contributed equally}
\affiliation{Racah Institute of Physics, Hebrew University of Jerusalem, Givat Ram, Jerusalem 91904 Israel}
\affiliation{QuTech and Kavli Institute of Nanoscience, Delft University of Technology, 2600 GA Delft, the Netherlands}
\author{D. LeBoeuf} 
\affiliation{Laboratoire National des Champs Magnétiques Intenses (LNCMI-EMFL), CNRS, UGA, UPS, INSA, Grenoble/Toulouse, France} 
\author{S. Ilić} 
\affiliation{Université Grenoble Alpes, CEA, Grenoble INP, IRIG, PHELIQS, 38000 Grenoble, France}  
\affiliation{Centro de Física de Materiales (CFM-MPC), Centro Mixto CSIC-UPV/EHU, 20018 Donostia-San Sebastián, Spain}
\author{M. Haim} 
\affiliation{Racah Institute of Physics, Hebrew University of Jerusalem, Givat Ram, Jerusalem 91904 Israel} 
\author{D.Möckli} 
\affiliation{Racah Institute of Physics, Hebrew University of Jerusalem, Givat Ram, Jerusalem 91904 Israel} 
\affiliation{Instituto de F\'{i}sica, Universidade Federal do Rio Grande do Sul, 91501-970 Porto Alegre, RS, Brazil}
\author{S. Kramer} 
\affiliation{Laboratoire National des Champs Magnétiques Intenses (LNCMI-EMFL), CNRS, UGA, UPS, INSA, Grenoble/Toulouse, France} 
\author{M. Khodas} 
\affiliation{Racah Institute of Physics, Hebrew University of Jerusalem, Givat Ram, Jerusalem 91904 Israel}
\author{M. Houzet} 
\affiliation{Université Grenoble Alpes, CEA, Grenoble INP, IRIG, PHELIQS, 38000 Grenoble, France}  
\author{J. S. Meyer} 
\affiliation{Université Grenoble Alpes, CEA, Grenoble INP, IRIG, PHELIQS, 38000 Grenoble, France}  
\author{M. Aprili} 
\affiliation{Laboratoire de Physique des Solides (CNRS UMR 8502), Bâtiment 510, Université Paris-Saclay 91405 Orsay, France} 
\author{H. Steinberg} 
\affiliation{Racah Institute of Physics, Hebrew University of Jerusalem, Givat Ram, Jerusalem 91904 Israel}
\author{C. H. L. Quay} 
\affiliation{Laboratoire de Physique des Solides (CNRS UMR 8502), Bâtiment 510, Université Paris-Saclay 91405 Orsay, France}

\begin{abstract}

In conventional Bardeen-Cooper-Scrieffer (BCS) superconductors, Cooper pairs of electrons of opposite spin (i.e. singlet structure) form the ground state. Equal spin triplet pairs (ESTPs), as in superfluid $^3$He, are of great interest for superconducting spintronics and topological superconductivity, yet remain elusive. Recently, odd-parity ESTPs were predicted to arise in (few-)monolayer superconducting NbSe$_2$, from the non-colinearity between the out-of-plane Ising spin-orbit field (due to the lack of inversion symmetry in monolayer NbSe$_2$) and an applied in-plane magnetic field. These ESTPs couple to the singlet order parameter at finite field. Using van der Waals tunnel junctions, we perform spectroscopy of superconducting NbSe$_2$ flakes, of 2--25 monolayer thickness, measuring the quasiparticle density of states (DOS) as a function of applied in-plane magnetic field up to 33T. In flakes $\lesssim$ 15 monolayers thick the DOS has a single superconducting gap. In these thin samples, the magnetic field acts primarily on the spin (vs orbital) degree of freedom of the electrons, and superconductivity is further protected by the Ising field. The superconducting energy gap, extracted from our tunnelling spectra, decreases as a function of the applied magnetic field. However, in bilayer NbSe$_2$, close to the critical field (up to 30T, much larger than the Pauli limit), superconductivity appears to be more robust than expected from Ising protection alone. Our data can be explained by the above-mentioned ESTPs. 

\end{abstract}


\maketitle  

\section{Introduction}

In both superfluid $^3$He and conventional Bardeen-Cooper-Schrieffer (BCS) superconductors, the ground state is made up of paired spinful entities, respectively nuclei and electrons. While the superfluid $^3$He wavefunction has a spin triplet structure, conventional superconductors are spin singlet~\cite{annett}.

The question thus arises of the possible existence of triplet superconducting pairs, and in particular equal spin triplet pairs (ESTPs, linear combinations of $\ket{\uparrow \uparrow}$ and $\ket{\downarrow \downarrow}$), as have been found in $^3$He-A~\cite{leggett}. ESTPs are intimately related to topological superconductivity and Majorana edge modes~\cite{frolov}. They are also of great interest for superconducting spintronics, as they can carry spin information without dissipation~\cite{ohnishi}.

\begin{figure}	
	\includegraphics[width=0.5\textwidth]{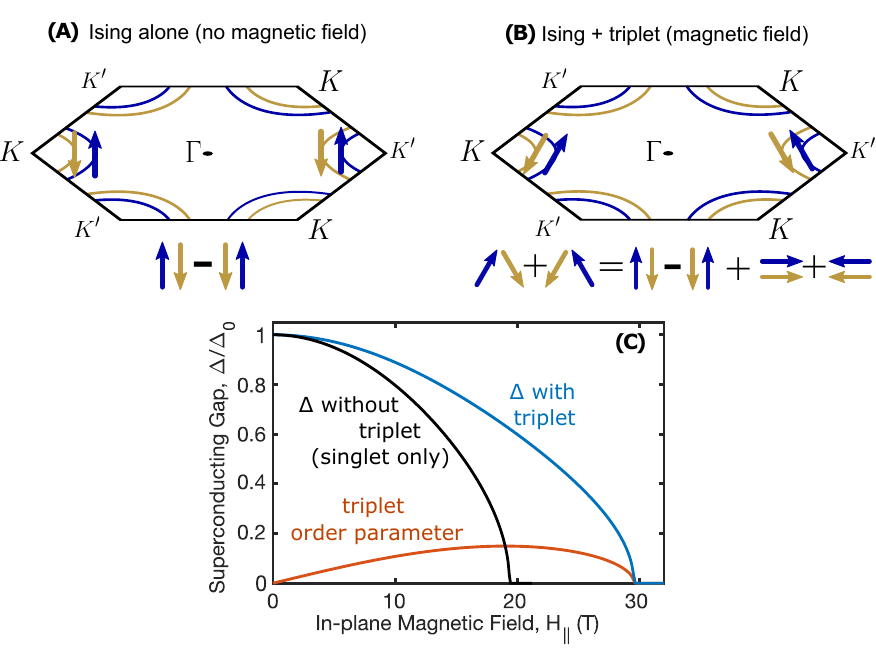}	
	\caption{(a) At zero magnetic field, singlet Cooper pairs are composed of electrons at opposite corners of the hexagonal NbSe$_2$ Brillouin zone (K and K' points). Their spins are pinned out-of-plane by the Ising field. (b) An in-plane magnetic field partially aligns electron spins orthogonal to the Ising field, and gives rise to odd-parity equal-spin triplet pairs~\cite{mockli2020}. (c) Theoretical superconducting gap as a function of in-plane magnetic field. Compared to the case where only the Ising field is considered (black line), superconductivity is even more robust to the in-plane magnetic field and the $\Delta$ vs $H_{||}$ curve has a `flattened' shape at intermediate fields (blue line). The triplet component of the order parameter (red line) with spin structure $\Phi_{tB}$  survives disorder through its coupling to the singlet component, which has spin structure $\Phi_{s}$. The temperature used in the calculations is 0.5 $T_c$, the critical temperature.
	See Figure~\ref{Figure3} and the text for details and comparison to data.
	}
	\label{Figure0}
\end{figure}

ESTPs have recently been predicted to arise in (few-)monolayer superconducting 2H-NbSe$_2$ (hereafter NbSe$_2$) in an applied in-plane magnetic field~\cite{mockli2020}, as follows: 

Monolayer transition metal dicalcogenides (TMDs), such as NbSe$_2$, with 2H structure lack in-plane crystal inversion symmetry; this gives rise, via the spin-orbit interaction, to an effective out-of-plane magnetic field $H_{SO}$ known as the `Ising (spin-orbit) field'~\cite{frigeri2004,gorkov2001,saito2016,lu2015}. $H_{SO}$ is momentum-dependent; in particular, it has opposite sign at K and K' points of the hexagonal Brillouin zone~\cite{xu2014}, and a predicted amplitude~\cite{wickramaratne2020} of $\mu_B H_{SO} = E_{SO} \approx 100$ meV in monolayer NbSe$_2$. As it is time-reversal invariant, $H_{SO}$ does not affect the strength of singlet superconductivity; however, it causes Cooper pair spins to point out-of-plane (Figure~\ref{Figure0}(a)) --- unlike conventional superconductors, where Cooper pairs' internal spin axes have no preferred direction.

Thus, an applied in-plane magnetic field $H_{||}$ never completely aligns Cooper pair spins in the plane even when the Zeeman energy $E_Z = \mu_{B}H_{||} \gg E_{SO}$: at zero temperature, the in-plane critical field $H_c$ is expected to diverge logarithmically~\cite{frigeri2004,ilic2017}. In agreement with these expectations, TMD superconductors of (few-)monolayer thicknesses obtained by exfoliation \cite{wang2012} or single-layer deposition ~\cite{zhang2010,fan2015} show critical fields $H_c$ much larger than $\mu_B H_P = \Delta_0/ \sqrt{g}$, the Pauli or paramagnetic limit~\cite{xi2016,liu2018}$^,$ \footnote{At the Pauli limit, the paramagnetic state of spin-aligned quasiparticles becomes more energetically favourable than the superconducting ground state~\cite{clogston1962, fulde1973}. In the absence of spin-orbit coupling, $H_P$ is thus the expected $H_c$ for thin superconductors, where the Meissner effect can be neglected.}. (Here $\mu_B$ is the Bohr magneton, $g$ the Landé g-factor and $\Delta_0$ the superconducting order parameter at zero field.) While inversion symmetry is restored in even-layered NbSe$_2$, the enhancement of $H_c$ persists in bilayer and few-monolayer TMDs, with $H_c$ decreasing monotonically with increasing NbSe$_2$ thickness~\cite{xi2016,barrera2018,prober1980}, and no observation of even-odd effects. This has been attributed to weak inter-layer coupling (compared to $E_{SO}$)~\cite{jones2014} and/or spin-layer locking~\cite{xi2016}.

Both singlet and opposite-spin triplet superconducting order parameters can exist at zero magnetic field, with spin structures respectively $ \Phi_s= \ket{\uparrow \downarrow} - \ket{\downarrow \uparrow}$ and $\Phi_t = \ket{\uparrow \downarrow} + \ket{\downarrow \uparrow}$~\cite{smidman2017}. For $E_{SO} < E_F$ (the Fermi energy), which is the case here, $ \Phi_t$ and $\Phi_s$ decouple, and $\Phi_t$ should not coexist with $\Phi_s$~\cite{rainer1998,haim2020}. ($\Phi_t$ is in any case sensitive to disorder and disappears when the mean free path is shorter than the superconducting coherence length~\cite{mockli2020}.)

The applied in-plane field $H_{||}$, due to its non-colinearity with the Ising field, as well as the momentum dependence of the latter, results in ESTPs with spin structure $\Phi_{tB}= \ket{\downarrow  \downarrow} + \ket{\uparrow \uparrow} $~\cite{mockli2020, tang2020, nakamura2017} (Figure~\ref{Figure0}(b)). $\Phi_{tB}$ is coupled to $\Phi_{s}$ by the in-plane field, and the critical field is affected by their symbiotic relationship~\cite{mockli2019}: $\Phi_{tB}$ enables $\Phi_{s}$ to survive the magnetic field, while $\Phi_{s}$ enables $\Phi_{tB}$ to survive disorder. As a result, in a disordered sample, or even when the temperature $T>T_{ct}$ (the critical temperature associated with $\Phi_{tB}$), the in-plane critical field is higher than it would be for either $\Phi_{s}$ or $\Phi_{tB}$ alone, and the dependence of the superconducting gap $\Delta$ on the applied field is also affected (Figure~\ref{Figure0}(c)).

Very recently, a two-fold anisotropy of critical field, non-linear transport and magneto-resistance was observed in few- and mono-layer NbSe$_2$ devices close to the transition to the normal state~\cite{Hamill_2020,Cho_Lortz_2020}. These results were also interpreted as coming from unconventional superconductivity: $\Phi_{tB}$ triplet components induced by the applied magnetic field and lateral lattice strain can reduce the six-fold rotational symmetry expected from the hexagonal crystal lattice to two-fold symmetry~\cite{Hamill_2020,Cho_Lortz_2020,haim2022}.

Here we report tunnelling spectroscopy of few-monolayer NbSe$_2$ devices over a wide range of applied in-plane magnetic fields, up to 30 T. As the magnetic field increases, our measurement of the superconducting gap $\Delta$ progressively deviates from the prediction based on pure singlet pairing. We find that this field-induced deviation can be explained by the onset of ESTPs in the form of $\Phi_{tB}$ (Figure~\ref{Figure0}).

\section{Results}

We consider a single-band superconductor, with hole pockets at the K/K' points, and include $\Phi_s$ and $ \Phi_{tB}$ correlations. As mentioned above, $\Phi_{tB}$ is coupled linearly to $\Phi_{s}$, and is expressed even when $\Delta_{tB}<\Delta{s}$. If we neglect inter-valley scattering, and if a finite pairing interaction is present in the $\Phi_{tB}$ channel as suggested by recent density functional theory (DFT) calculations~\cite{wickramaratne2020}, the superconducting energy gap $\Delta$ can be obtained from the quasiclassical theory of superconductivity (cf. Equation S14 in Supp. Info.) :

\begin{equation} \label{singlet-triplet}
\Delta= (E_{SO}\Delta_{s}+ E_Z  \Delta_{tB})/\sqrt{E_{SO}^2+E_{Z}^2},
\end{equation} 

\noindent where $\Delta_{s}$ and $\Delta_{tB}$ are, respectively, the singlet and equal-spin triplet order parameters. 

Here we can see that, compared to the case of $\Phi_{s}$ with Ising protection alone, the coexistence of $\Delta_{tB}$  with $\Delta_{s}$ and the coupling between the two can increase the robustness of superconductivity against an applied in-plane magnetic field. In the case where there is no pairing in the equal-spin triplet channel ($\Delta_{tB} = 0$), $\Delta$ is reduced by the magnetic field through the factor $E_{SO}/\sqrt{E_{SO}^2+E_{z}^2}$ and vanishes asymptotically, giving the afore-mentioned logarithmic divergence of the critical field at zero temperature. To obtain the order parameters $\Delta_{s}$ and $\Delta_{tB}$ at finite temperature and magnetic field, one has to solve two coupled equations self-consistently (cf. Supp. Info. II). 

The quasiclassical theory also gives the density of states (DOS), which is found for $E<E_{SO}$ to be simply the BCS DOS, with the gap as in Equation~\ref{singlet-triplet} (see Equation 14 in the Supp. Info.) --- unlike 2D superconductors with low spin-orbit coupling in in-plane fields, the coherence peak is not Zeeman-split~\cite{meservey1970}. In addition, Ising protection gives a sharp coherence peak in the DOS, regardless of the strength of the triplet coupling or the applied magnetic field. Nevertheless, in the presence of inter-valley scattering ($\tau_{iv}$ being the inter-valley scattering time), Ising protection is reduced (due to averaging over valleys with opposite signs of $H_{SO}$), the DOS is smeared out \cite{haim2020} as in the Abrikosov-Gor'kov theory~\cite{abrikosov,bruno1973}, and the divergence of the critical field at zero temperature is regularised~\cite{ilic2017}. In the limit of strong inter-valley scattering ($E_{SO}^2/\Delta_s \gg 1/\tau_{iv} \gg \Delta_0$) the dependence of $\Delta$ on the applied magnetic field becomes similar to that expected from the Abrikosov-Gor'kov theory with the critical field given by  $\mu_B H_c = E_{SO} \sqrt{2 \Delta \tau_{iv} /\hbar}$. In our experiment, we do not have strong inter-valley scattering as $1/\tau_{iv} \lesssim \Delta_s$. (See Supp. Info. IC)

 \begin{figure}	
	\includegraphics[width=0.75\textwidth]{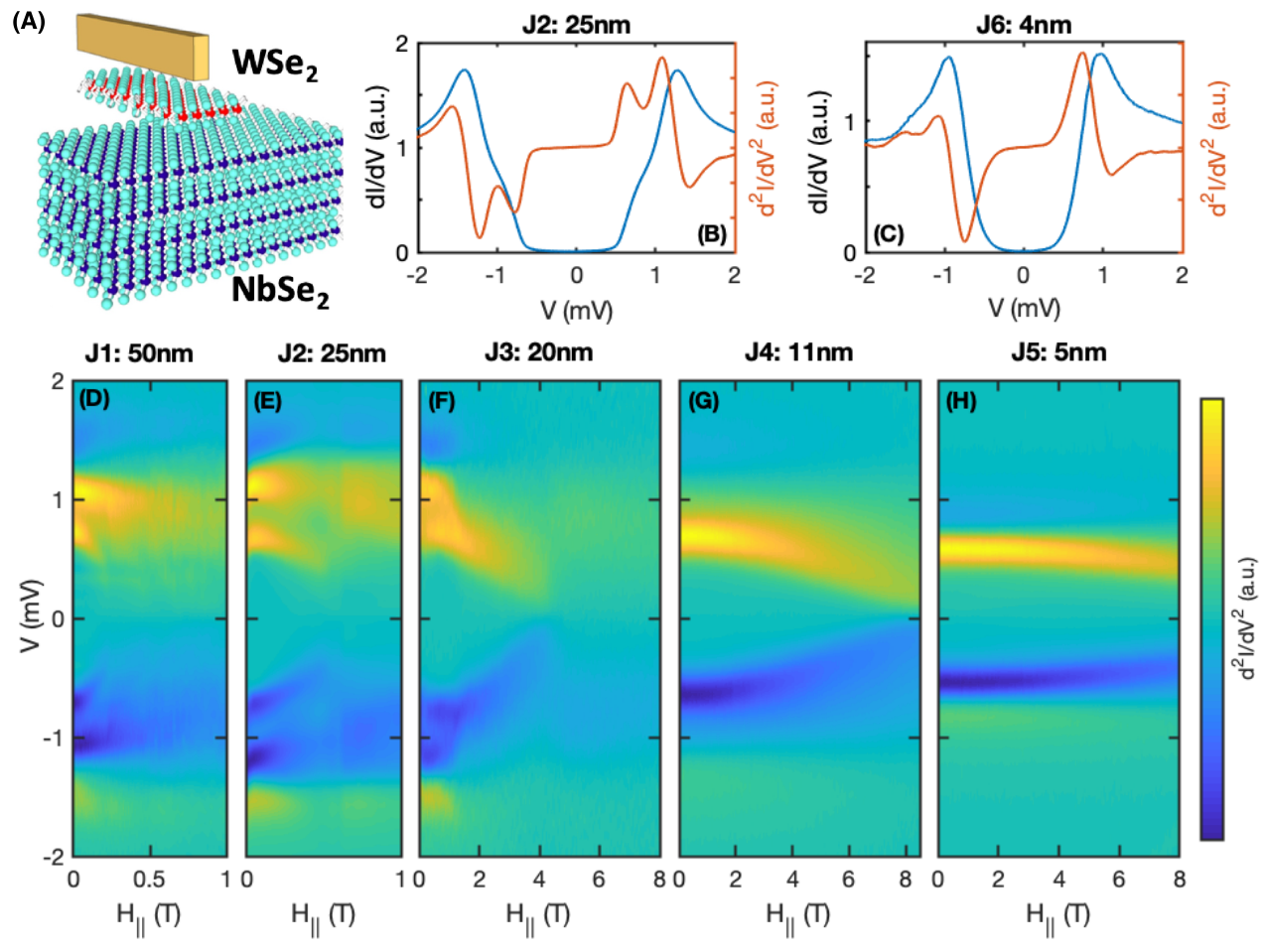}	
	\caption{Tunnelling spectroscopy of bulk and few-monolayer NbSe$_2$ through van der Waals barriers. (a) Schematic drawing of the tunnel junction: few-monolayer NbSe$_2$, covered with few-monolayer WSe$_2$ (or MoS$_2$) and a Ti/Au electrode. A voltage $V$ is applied between the Ti/Au electrode and the NbSe$_2$ and a current $I$ measured. (b) Differential conductance ($G = dI/dV$) as a function of $V$ across J2 (blue) and $d^2I/dV^2$ vs $V$ of J2 (red). A double peak/dip can be seen in $d^2I/dV^2$, due to the presence of two superconducting gaps. (c) Same as panel (b) for J6. (d)-(h) Colormaps of the magnetic field dependence of the $d^2I/dV^2$ curves for junctions J1--J5. The double dip/peak feature (yellow/blue regions) disappears in thin samples; a single gap is left. Measurements were taken at temperatures of 30-70 mK. }
	\label{Figure1}
\end{figure}

We fabricate tunnel junctions (J1 to J7) on superconducting NbSe$_2$ flakes of $1.2-50$ nm thickness. The tunnel barriers are thin flakes of semiconducting $\mathrm{WSe_2}$ or $\mathrm{MoS_2}$ exfoliated by the van der Waals dry transfer technique described in Ref.~\cite{dvir}. A Ti/Au normal counter electrode is then evaporated on the semiconductor leading to a structure shown schematically in Figure~\ref{Figure1}(a). An ohmic contact to the NbSe$_2$ is also fabricated. The typical surface area of the junction is about 1~µm$^2$ and the resistance in the normal state $>$10~k$\Omega$. The critical temperature $T_c$ decreases from $\sim$7.2 K in the thickest flakes to $\sim$2.6 K in the thinnest ones. 

Using standard lock-in techniques, we first measure the current $I$ and differential conductance $G=dI/dV$ across the junctions as a function of applied bias voltage $V$~\cite{giaever1960} and in-plane magnetic fields $H_{||}$ in dilution refrigerators with base temperatures of 30--70 mK. $G(V)$ is proportional to the DOS convolved with the derivative of the Fermi distribution function \cite{tinkham}. Therefore, in principle, the energy resolution of our spectroscopy is given by the temperature and the integrated voltage noise across the junction. 

Typical $G(V)$ curves are shown for a 25 nm thick sample (J2) and a 6 monolayer sample (J6) in Figure~\ref{Figure1}(b,c), top panels. The main differences between these junctions are: (1) the smaller superconducting gap in the thinner device due to a smaller $T_c$, and (2) the low-energy shoulder, very clearly seen in the thicker junction, is absent in the thinner one. This is even more apparent in the second derivative of the current as a function of the voltage bias, $dG/dV$, in Figure~\ref{Figure1}(b,c): the two peaks in J2, merge to a single peak in J6. This merging was previously observed \cite{dvir, khestanova2018} and we now see that it persists in flakes up to 11 nm ($\approx 15$ monolayer) thick: the two-gap superconductivity of bulk NbSe$_2$~\cite{noat2015} is lost. This is consistent with band structure calculations for bulk and monolayer NbSe$_2$: whereas in the bulk three bands cross the Fermi level~\cite{johannes,yokoya}, and two superconducting gaps have been observed~\cite{dvir}, in the monolayer a single band remains, which crosses the Fermi level twice, resulting in hole pockets at the K/K' and $\Gamma$ points~\cite{wickramaratne2020}. A single-band theory thus seems most suitable for the thinnest flakes. 

Figures~\ref{Figure1}(d-h) show the evolution of the $dG/dV$ curves of five junctions (J1-J5) with increasing in-plane magnetic field. Junctions 1 and 2, the thickest, show similar responses to the applied field: the inner peak shifts to lower energies faster than the outer peak. This is consistent with previous experiments, and is likely due to the 3D character of the Se $p_z$-orbital-derived band at the $\Gamma$ point, which is associated with the smaller superconducting energy gap, as well as its higher diffusion coefficient~\cite{dvir,dvir_nanoletters}. For the thinner junctions, J4 and J5, a single gap persists from zero field up to 9 T.

As noted above, the robustness of the gap to applied magnetic fields is expected in thin samples due to Ising protection and drastically reduced orbital depairing (Meissner effect). To significantly reduce the gap and to study the effect of the applied field on the density of states it is necessary to go to even higher fields. 

 \begin{figure}[h!]
	\includegraphics[width=0.5\textwidth]{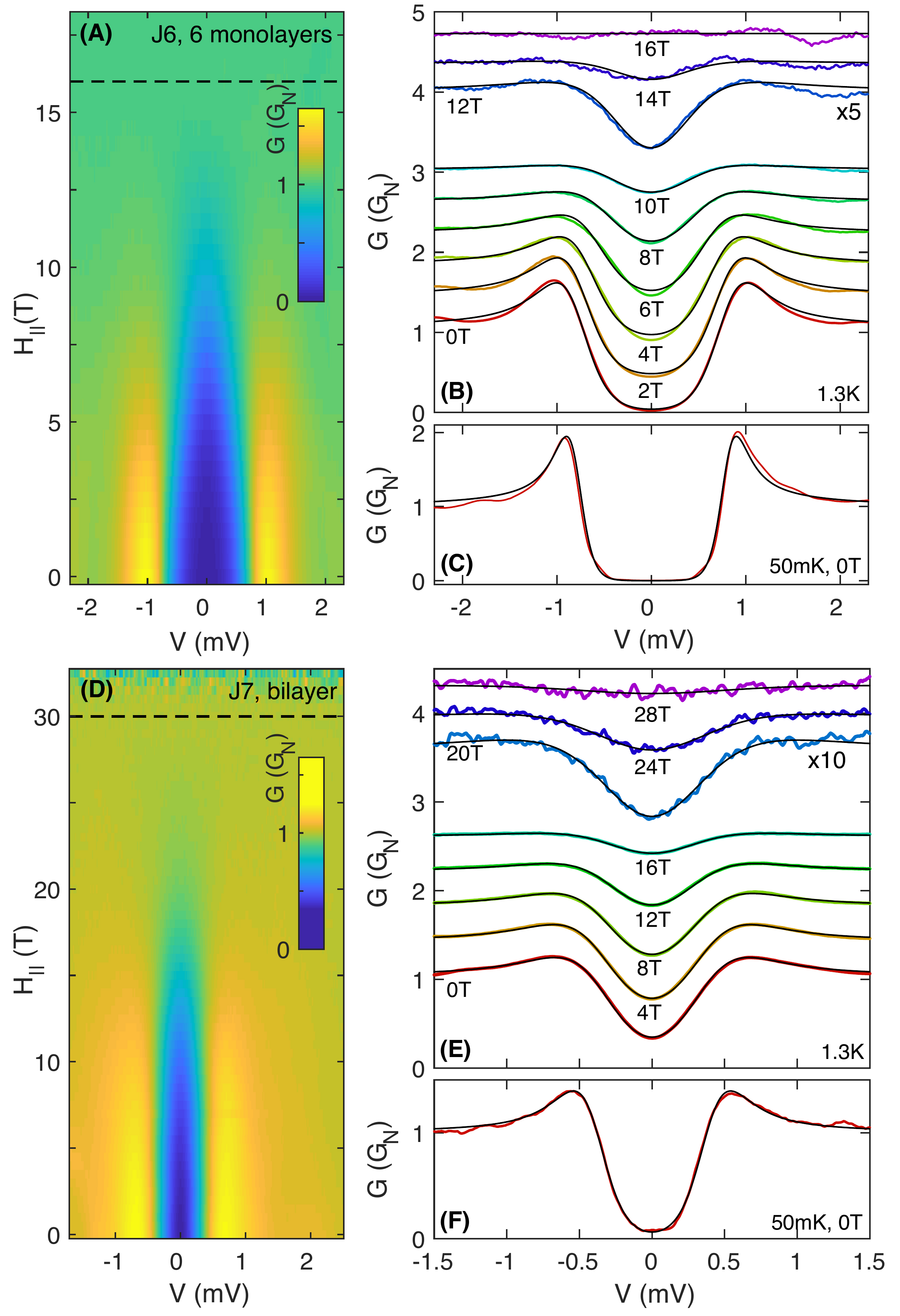}	 
	\caption{Differential conductance $G = dI/dV$ as a function of the bias voltage $V$ and of the in-plane magnetic field $H_{||}$ of J6 (6 monolayers, top panels) and J7 (bilayer, bottom panels). The tunnelling spectra are normalized by the normal state conductance, $G_N(V)$, measured above $H_c$. (a,d) Colormap of $G(V)$ as a function of field at 1.3 K. The dotted lines indicate the critical fields. (b,e) Horizontal slices of the data in the colormaps (a) and (d) respectively, showing $G(V)$ at different fields, vertically displaced for clarity. The black lines are fits to an Abrikosov-Gor'kov-like (A-G-like) density of states, with the energy gap and A-G broadening parameter as fitting parameters. (The gap is not determined self-consistently.) (c,f) Data at 50mK and zero magnetic field (red lines) together with the fits obtained using a BCS DOS and an effective temperature (black lines). The superconducting gaps obtained from the fits are, respectively, 800$\mu$eV and 400$\mu$eV, while the effective temperatures are respectively 0.9K and 1K.}
	\label{Figure2}
\end{figure}

Therefore, we measure two tunnel junctions (J6, 6 monolayers) and (J7, bilayer) in in-plane magnetic fields of up to 33 T at 1.3 K (pumped liquid helium). Their critical temperatures are, respectively, 5.4 K ($H_P = 10.5$ T) and 2.6 K ($H_P = 5$ T), giving $\Delta/k_B T_c\approx 1.8$, close to the BCS prediction and in agreement with previous studies~\cite{dvir,khestanova2018}. (See Figure S2 and inset.) Finally, the critical in-plane fields are $H_c$ = 18 T for J6 and $H_c$ = 30 T for J7, corresponding respectively to $H_c=1.5 H_P$ and $H_c=6 H_P$. (See Figure~\ref{Figure3}.) These junctions had earlier been characterized at 50mK (dilution refrigerator) at zero magnetic field (Figures~\ref{Figure2}(c) and (f)) --- hard gaps were observed, pointing to tunnelling as the main transport mechanism. These tunnel spectra are well-described by a fit to a BCS density of states, broadened by a $\sim 200\mu eV$ effective temperature. Though higher than the bath temperature, this broadening does not affect the determination of the energy gap, which can be done with high precision. (See Supp. Info. IA and IB for details.)

\section{Discussion}

The evolution of $G(V)$ with the in-plane magnetic field at 1.3K is shown in Figures~\ref{Figure2}(a) (J6) and ~\ref{Figure2}(d) (J7). For clarity, spectra at selected magnetic fields are also shown in Figures~\ref{Figure2}(b) and 2(e) together with an Abrikosov-Gor'kov-like density of states with a field-dependent broadening parameter\cite{abrikosov,srivastava2008}, convolved with a Fermi function to account for the temperature. The fits account very well for the experimental data. 

\begin{figure}[h!]
	\includegraphics[width=0.5\textwidth]{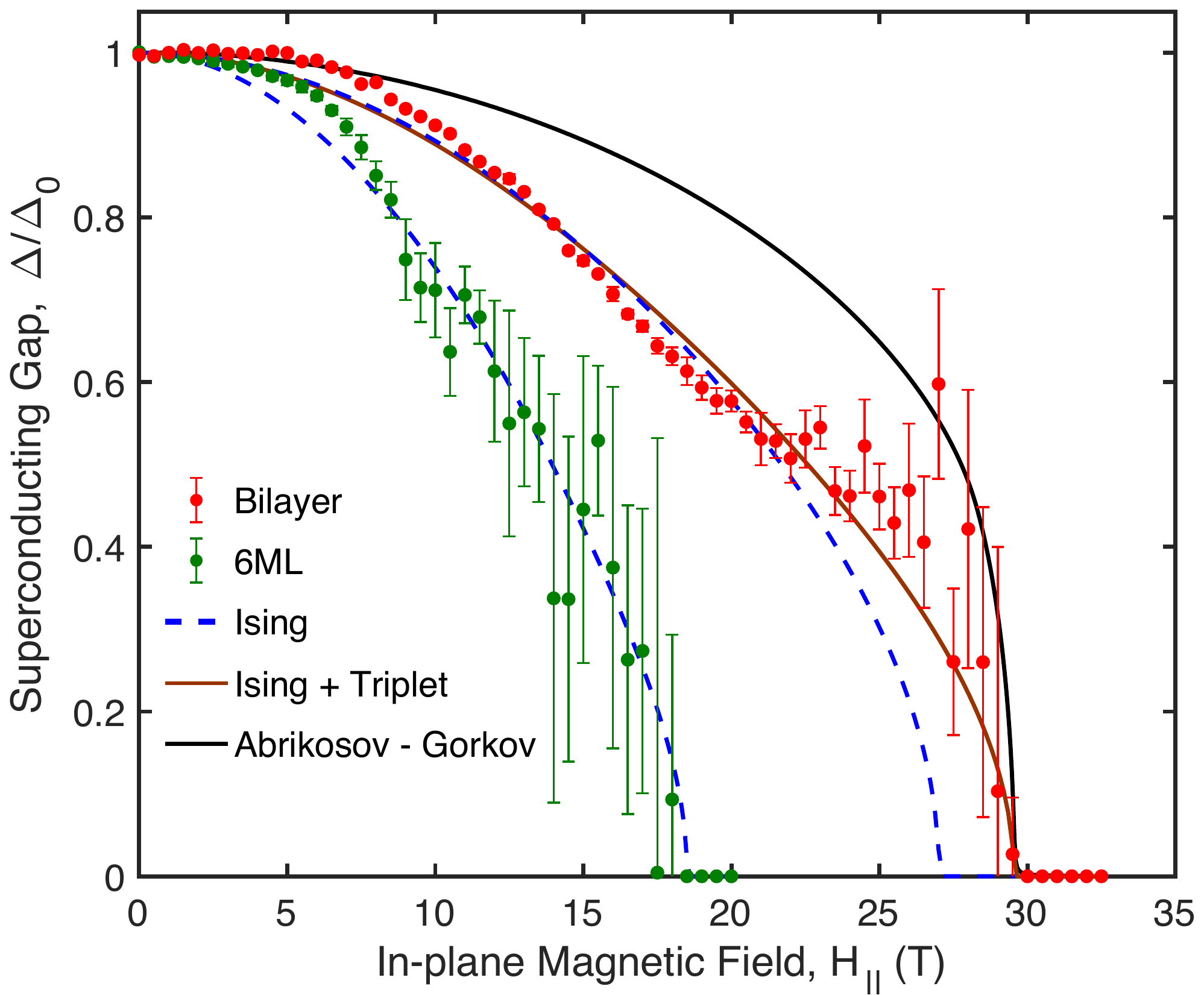}	
	\caption{Normalized superconducting gap as a function of the in-plane magnetic field $\Delta(H_{||})$ obtained from the fits of the quasiparticle density of states in Figure~\ref{Figure2}. The error bars were obtained following the procedure described in Supp. Info. IA. The dark blue dashed lines are a fit of experimental data using the Ising theory alone. Here, $E_{SO}=14.45 T_{cs}$ (with $T_{cs}=2.6K$) for the bilayer and $E_{SO}=2.21 T_{cs}$ (with $T_{cs}=5.4K$) for the 6-monolayer. In brown is the Ising theory with an equal-spin triplet component of the order parameter as described in the text. Here, $E_{SO}=9.62 T_{cs}$ and $T_{ct}=0.05 T_{cs}$, with $T_{cs}=2.6K$. 
	Finally, the solid black line is calculated using the Ising theory with strong disorder (equivalent to the Abrikosov-Gor'kov theory). In all cases, the critical field is constrained to be the experimental one.}
	\label{Figure3}
\end{figure}

The superconducting gaps obtained from these fits are shown as a function of the in-plane magnetic field in Figure~\ref{Figure3}, and compared to theory. 

For the six-monolayer device, a simple Ising model accounts for the data reasonably well (Figure~\ref{Figure3}, dashed dark blue line). The fitting parameters are given in the caption of the figure. 

Focusing on the thinner, bilayer device (J7), we see that the Ising theory alone without triplet pairing 
fits the data reasonably well up to about 20T, but not close to the critical field, where the superconducting energy gap is more robust than expected (Figure~\ref{Figure3}, dashed dark blue line). 

This key experimental finding is suggestive of a second order parameter, which is revealed as the dominant order parameter disappears \cite{rainer1998}. Indeed introducing a small ESTP component of the gap (triplet model), a better fit of the overall experimental data is obtained (Figure~\ref{Figure3}, brown line). The temperature of the experiment (1.3K) is above the triplet critical temperature ($T_{ct}=0.05T_{cs}=$130mK, obtained from the fit). Therefore, the ESTP order parameter $\Delta_{tB}$ exists only through its coupling with the singlet order parameter $\Delta_s$, and its main effect is to enhance the critical field through the coupling with the singlet order parameter. In addition, the triplet subdominant component also renders the gap vs field dependence more linear (Figure~\ref{Figure0}(c)).

Our fit also gives $E_{SO}$ = $9.62 T_{cs}$ ($\sim$ 2.2 meV). This is a lower bound for $E_{SO}$, as inter-valley (K-K') scattering is not taken into account. If it is, higher values of $E_{SO}$ will have to be used to arrive at the same $H_{||}^c$, but the shape of the $\Delta(H_{||})$ curve is similar. Further, we note that the shape of $\Delta(H_{||})$ in the triplet model is by construction impervious to intra-valley scattering. Our $E_{SO}$ value is consistent with the upper bound for $E_{SO}$ given by angle-resolved photo-emission spectroscopy (ARPES) measurements, which indicate $E_{SO} \lesssim 20$meV (the measurement resolution), significantly lower than theoretical predictions~\cite{xu2018}. 

For completeness, we also show the Ising theory with strong inter-valley scattering (equivalent to Abrikosov-Gor'kov), where the only fitting parameter is the critical field (Figure~\ref{Figure3}, black line). This does not fit the data at all -- the experimental $\Delta$ is consistently smaller than that predicted by the theory, which also fails to reproduce the `linear' part of the curve at intermediate fields.

At present, much of the literature on quantum transport in few-layer NbSe$_2$ includes only the hole pockets at the K/K' points, as we did, even though there is also a hole pocket at the $\Gamma$ point~\cite{wickramaratne2020}. In Supp. Info. IIC and IID, we consider models which include only $\Phi_s$ and K/K'-$\Gamma$ coupling, and neglect all triplet order parameters. These are found not to fit our data well, given the known level of disorder in the sample, thus strengthening the case for the ESTP interpretation. 

Regarding ESTPs, we note that, within the scenarios of Refs.~\cite{Hamill_2020,Cho_Lortz_2020} mentioned earlier, the triplet order parameters allowed by symmetry such as $\Phi_{tB}$ have to be nearly degenerate with the leading singlet order parameter. Attraction in the triplet channel is supported by recent density functional theory (DFT) calculations \cite{wickramaratne2020}; however, there is at present no evidence of near degeneracy between triplet and singlet channels. Our interpretation does not require near degeneracy, and the singlet-triplet coupling comes from a clear microscopic mechanism (the in-plane magnetic field), which is quantitatively accounted for both in the theory and in the analysis of the experimental data.

While previous reports on Andreev spectroscopy experiments have shown a reduction of the gap consistent with field-induced depairing in the presence of Ising protection~\cite{sohn2018}, our hard-gap tunnel junctions allow a nuanced and quantitative analysis of the possible microscopic mechanisms for the enhancement of the critical field, pointing to the presence of equal-spin triplet superconductivity.

Further study at even lower temperatures, independent measurements of $E_{SO}$, independent estimates of the K/K'-$\Gamma$ coupling from theory or experiment, and momentum-selective barriers would be helpful to unambiguously confirm the existence of ESTPs in NbSe$_2$.

\section{Materials and Methods} Especially at high magnetic fields, special care was taken to ensure that the applied magnetic field is parallel to the flakes. It is aligned to better than $\sim$1$^\circ$. In addition, we checked that the voltage noise due to mechanical vibrations is lower than that from the thermal broadening. This is described in detail in Supp. Info. IA.

\section{Data Availability} The datasets generated and/or analysed during the current study are available from the corresponding author upon reasonable request.
 
\section{Acknowledgements} We acknowledge valuable discussions with Pascal Simon and Freek Massee, and thank the latter for a careful reading of the manuscript. This work was funded by a Maimonides-Israel grant from the Israeli-French High Council for Scientific and Technological Research; JCJC (SPINOES), PIRE (HYBRID) and PRC (TRIPRES) grants from the French Agence Nationale de Recherche; ERC Starting Grant ERC-2014-STG 637928 (TUNNEL); Israel Science Foundation Grants No.s 861/19 and 2665/20, and the Laboratoire d’Excellence LANEF in Grenoble (ANR10-LABX-51-01). T.D. is grateful to the Azrieli Foundation for an Azrieli Fellowship. Part of this work has been performed at the Laboratoire National de Champs Magnétiques Intenses (LNCMI), a member of the European Magnetic Field Laboratry (EMFL).

\bibliographystyle{naturemag}

\bibliography{references}

\end{document}


\widetext

\preprint{APS/123-QED}

\title {\large SUPPLEMENTAL MATERIAL for\\[0.25cm] \lq\lq Tunnelling spectroscopy of few-monolayer NbSe$_2$ in high magnetic field: triplet superconductivity and Ising protection\rq\rq}
\author{M. Kuzmanović}
\thanks{These two authors contributed equally}
\affiliation{Laboratoire de Physique des Solides (CNRS UMR 8502), Bâtiment 510, Université Paris-Saclay 91405 Orsay, France}  
\affiliation{QTF Centre of Excellence, Department of Applied Physics, Aalto University School of Science, P.O. Box 15100, 00076 Aalto, Finland}
\author{T. Dvir} 
\thanks{These two authors contributed equally}
\affiliation{Racah Institute of Physics, Hebrew University of Jerusalem, Givat Ram, Jerusalem 91904 Israel}
\affiliation{QuTech and Kavli Institute of Nanoscience, Delft University of Technology, 2600 GA Delft, the Netherlands}
\author{D. LeBoeuf} 
\affiliation{Laboratoire National des Champs Magnétiques Intenses (LNCMI-EMFL), CNRS, UGA, UPS, INSA, Grenoble/Toulouse, France} 
\author{S. Ilić} 
\affiliation{Université Grenoble Alpes, CEA, Grenoble INP, IRIG, PHELIQS, 38000 Grenoble, France}  
\affiliation{Centro de Física de Materiales (CFM-MPC), Centro Mixto CSIC-UPV/EHU, 20018 Donostia-San Sebastián, Spain}
\author{M. Haim} 
\affiliation{Racah Institute of Physics, Hebrew University of Jerusalem, Givat Ram, Jerusalem 91904 Israel} 
\author{D.Möckli} 
\affiliation{Racah Institute of Physics, Hebrew University of Jerusalem, Givat Ram, Jerusalem 91904 Israel} 
\affiliation{Instituto de F\'{i}sica, Universidade Federal do Rio Grande do Sul, 91501-970 Porto Alegre, RS, Brazil}
\author{S. Kramer} 
\affiliation{Laboratoire National des Champs Magnétiques Intenses (LNCMI-EMFL), CNRS, UGA, UPS, INSA, Grenoble/Toulouse, France} 
\author{M. Khodas} 
\affiliation{Racah Institute of Physics, Hebrew University of Jerusalem, Givat Ram, Jerusalem 91904 Israel}
\author{M. Houzet} 
\affiliation{Université Grenoble Alpes, CEA, Grenoble INP, IRIG, PHELIQS, 38000 Grenoble, France}  
\author{J. S. Meyer} 
\affiliation{Université Grenoble Alpes, CEA, Grenoble INP, IRIG, PHELIQS, 38000 Grenoble, France}  
\author{M. Aprili} 
\affiliation{Laboratoire de Physique des Solides (CNRS UMR 8502), Bâtiment 510, Université Paris-Saclay 91405 Orsay, France} 
\author{H. Steinberg} 
\affiliation{Racah Institute of Physics, Hebrew University of Jerusalem, Givat Ram, Jerusalem 91904 Israel}
\author{C. H. L. Quay} 
\affiliation{Laboratoire de Physique des Solides (CNRS UMR 8502), Bâtiment 510, Université Paris-Saclay 91405 Orsay, France}

\maketitle 

\renewcommand{\theequation}{S\arabic{equation}}
\renewcommand{\thefigure}{S\arabic{figure}}
\renewcommand{\bibnumfmt}[1]{[S#1]}
\renewcommand{\citenumfont}[1]{S#1}

The supplementary material is divided into two parts. Part \ref{partI} covers the experimental methods: section \ref{partI:GV} details how the order parameter was extracted from the spectroscopic data and gives an estimate of the uncertainty, while sections \ref{partI:DOSbroadening} and \ref{partI:scattering} comment on the broadening of the density of states and the disorder respectively. Part \ref{partII} presents the details of the theoretical model, where a general overview is given in section \ref{partII:hamiltonian}, followed by the triplet model in section \ref{sec-KK'}. At the end two-pocket models are discussed: 
The possibility of a Suhl-Matthias-Walker coupling is discussed in section \ref{sec:two-pockets-MSW}, and is ruled out on the basis of disorder, and a McMillan coupling is discussed in section \ref{sec-mm}, which is ruled out based on spectroscopic and transport data.

\bigskip

\section{The sample and experimental methods}\label{partI}

\begin{figure}[h]
	\centering
	\includegraphics[width=0.5\textwidth]{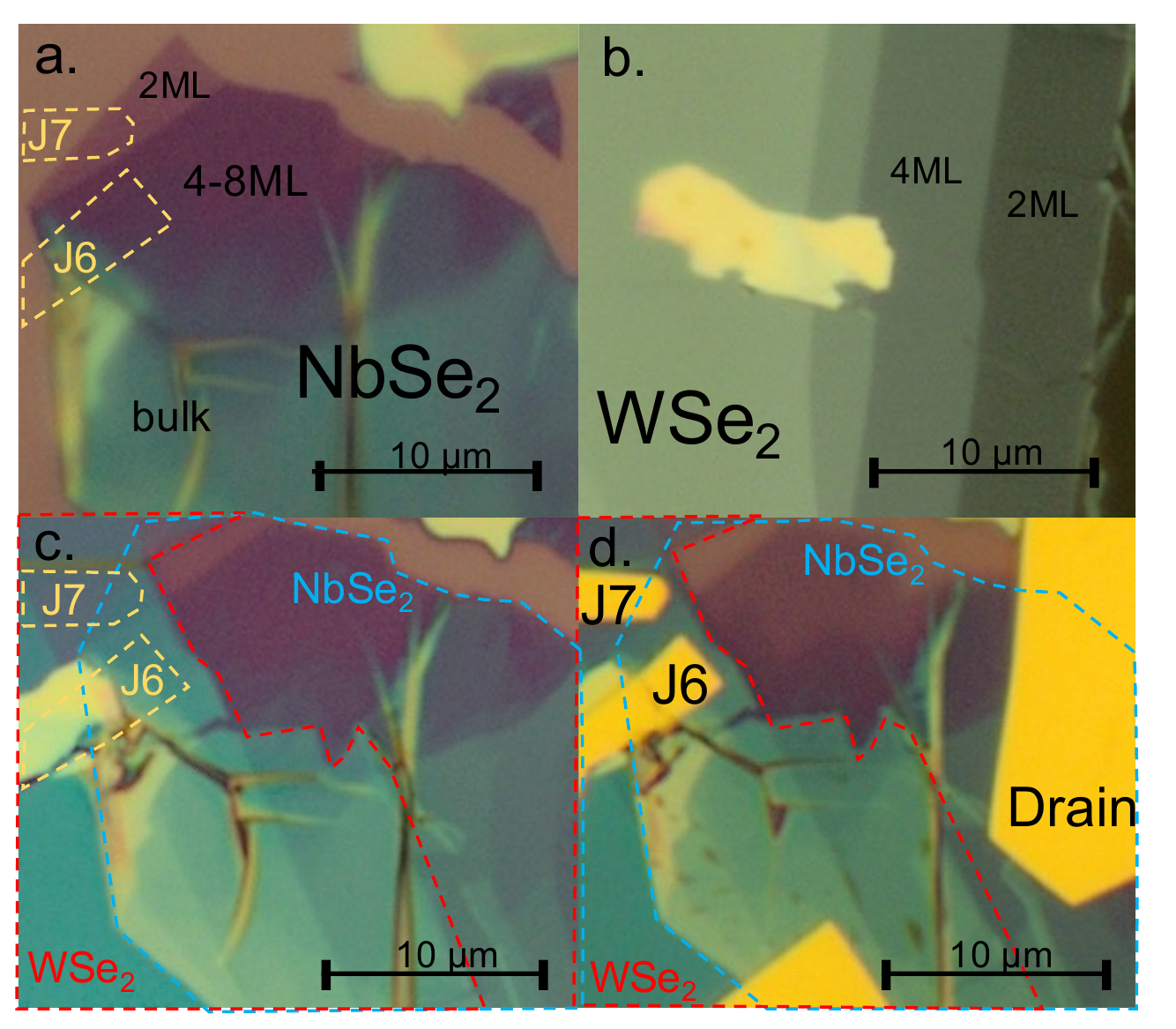}
	\caption{
		a. $\mathrm{NbSe_2}$ exfoliated on $\mathrm{SiO_2}$ from a PDMS stamp. Flake thickness is
		determined from the optical contrast. b. $\mathrm{WSe_2}$ exfoliated on a PDMS stamp.
		Flake thickness is determined from the optical contrast. c. $\mathrm{NbSe_2 - WSe_2}$
		heterostructure formed by the deterministic transfer of the WSe 2 . d. The
		final device with multiple tunnel junctions (J6,J7) and ohmic contacts
		(Drain).
	}
	\label{fig:sample}
\end{figure}

The devices reported in this work were fabricated in a similar method to those reported in ref \cite{dvir2018}. First, $\mathrm{NbSe_2}$ was exfoliated within a glovebox with an inert N2 environment onto a silicon chip covered with 285nm of $\mathrm{SiO_2}$. Next, $\mathrm{WSe_2}$ was exfoliated on a PDMS stamp and examined to look for thin flakes. Once a suitable flake was found, it was aligned and brought into contact with a thin flake of $\mathrm{NbSe_2}$. Next, the heterostructure was removed from the glovebox and tunnel contacts (5nm/80nm Ti/Au) were deposited on the $\mathrm{WSe_2}$ barrier. Finally, ohmic contacts (5nm/80nm Ti/Au) where deposited directly on the $\mathrm{NbSe_2}$ after Ar milling to remove oxide layers. The resulting device is shown in figure \ref{fig:sample}. The two junctions focused on in the main text are J7, on top of a 2ML $\mathrm{NbSe_2}$ flake, with a tunneling resistance of $R_1\approx 1.8\mathrm{M\Omega}$, and J6, on top of a 4-8ML $\mathrm{NbSe_2}$ region,  with a resistance of $R_2\approx 12\mathrm{k\Omega}$.

\begin{figure}[h]
	\centering
	\includegraphics[width=0.5\textwidth]{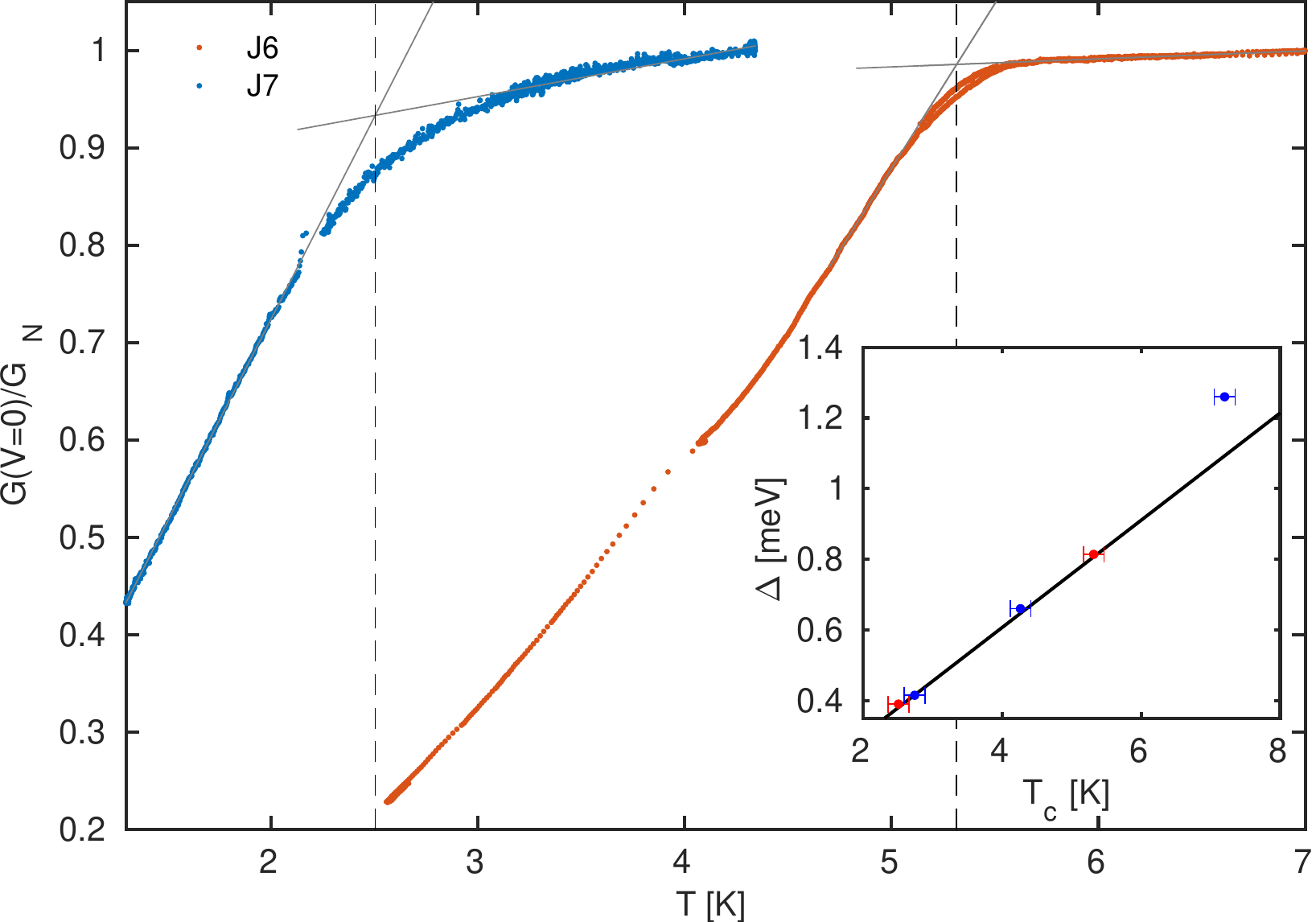}
	\caption{The temperature dependence of the zero voltage conductance, $G(V=0)$, for devices J7 (2ML) and J6 (6ML).
		Inset the $\Delta$ vs the estimated $T_c$: for J7 and J6 (red dots), the data from \cite{dvir2018spectroscopy} (blue circles) and the BCS result $\Delta=1.76k_bT_c$ (black line).
	}
	\label{fig:GvsT}
\end{figure}

As a part of the sample characterization the differential conductance of both junctions was measured at $T=50\mathrm{mK}$ (in a $\mathrm{He^3-He^4}$ dilution refrigerator) (shown in figure 3 of the main text). The extracted values for $\Delta$ are in line with the critical temperature $T_c$, estimated from the $G(V=0)$ temperature dependence - see figure \ref{fig:GvsT}.
This, as well as the data reported in \cite{khestanova2018unusual}, shows that for few-layer samples, which are of main interest here, the order parameter $\Delta$ and the critical temperature $T_c$ follow the BCS prediction $\Delta= 1.76k_bT_c$.

The high field measurements were performed at the "Laboratoire National des Champs Magnétiques Intenses" in Grenoble, France. The magnet used allowed for magnetic fields up to $H=36\mathrm{T}$.

\begin{figure}[h]
	\centering
	\includegraphics[width=0.7\textwidth]{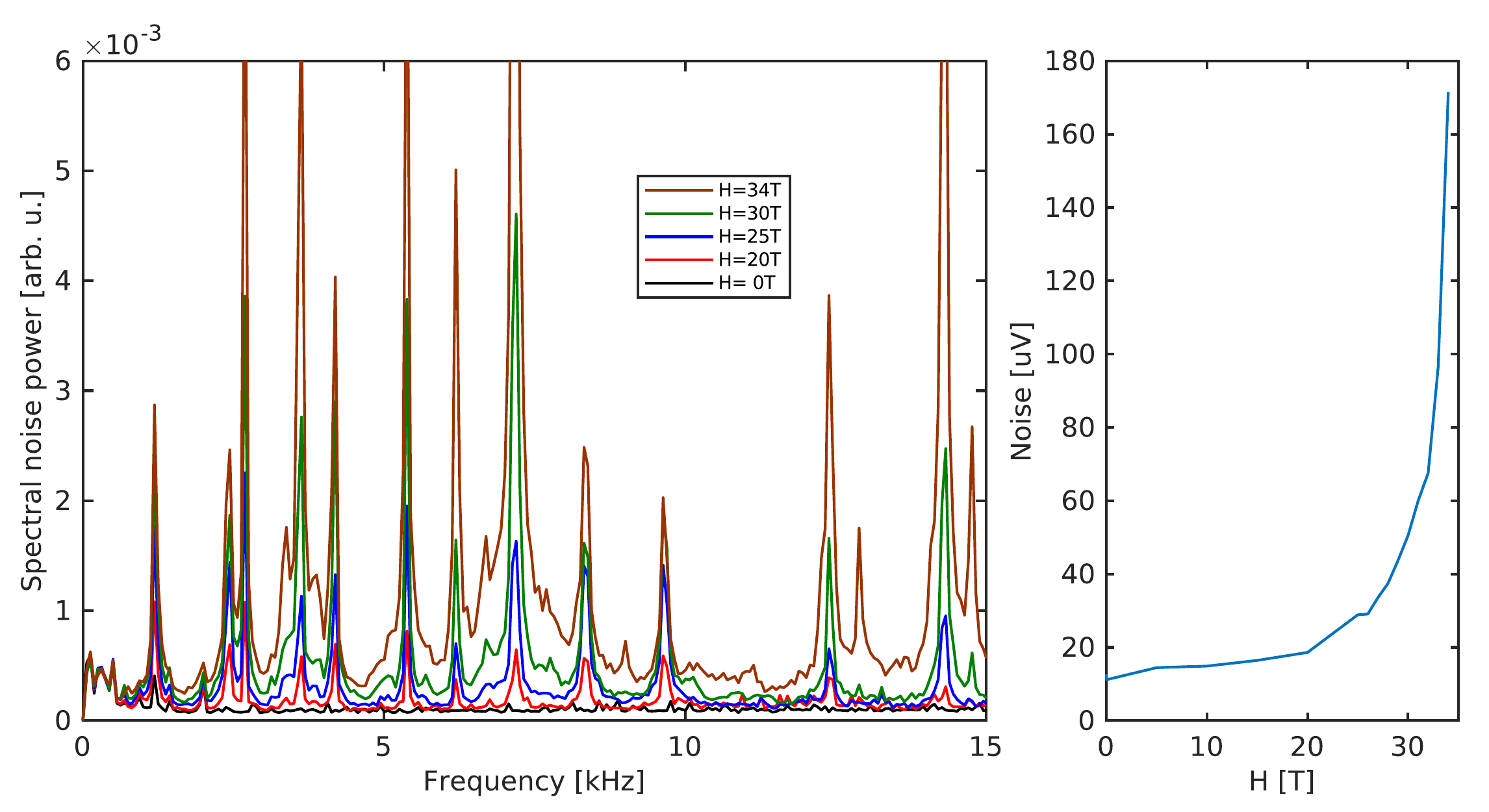}
	\caption{Electrical noise measured at $T=1.2\mathrm{K}$, across  a $R=10\mathrm{k \Omega}$ resistor in a bandwidth of $\mathrm{DC}-32\mathrm{kHz}$, as a function of the magnetic field. The left panel show the noise spectrum, while the right one shows the integrated noise voltage.
	}
	\label{fig:noise}
\end{figure}

The sample was cooled down to $T=1.25\mathrm{K}$ in a pumped $\mathrm{He^4}$ cryostat, and the differential conductance was measured using a standard lock-in technique with an excitation of $V_{lock-in}=100\mathrm{\mu V}$. Although the excitation voltage is relatively large, it does not lead to a significant smearing of the obtained spectrum as $eV_{lock-in}\ll3.5k_B T \approx 360\mu eV$ (where $3.5k_b T$ is the FWHM of the Fermi distribution transition). Likewise the electrical noise, generated by the magnet flux noise is also smaller than the temperature - see figure \ref{fig:noise}. The sample was mounted on a rotating stage, the axis of which was perpendicular to the magnetic field. A field of $H=2\mathrm{T}$ was applied and the sample was rotated, while measuring the height of the coherence peaks. By maximizing the peak height the magnetic field was aligned with the plane of the sample, with a maximum deviation of up to $1^\circ$. A misalignment of $\sim$1$^\circ$ (the maximum realistically  possible in our experiment), gives a perpendicular component of the magnetic field smaller than 0.3 T, several times smaller than the perpendicular critical field \cite{tsen2016,haim2020signatures}, thus it should not substantially decrease the observed critical field $H_c$.

\subsection{$G(V)$ curve fitting}\label{partI:GV}

\begin{figure}[h]
	\centering
	\includegraphics[width=0.5\textwidth]{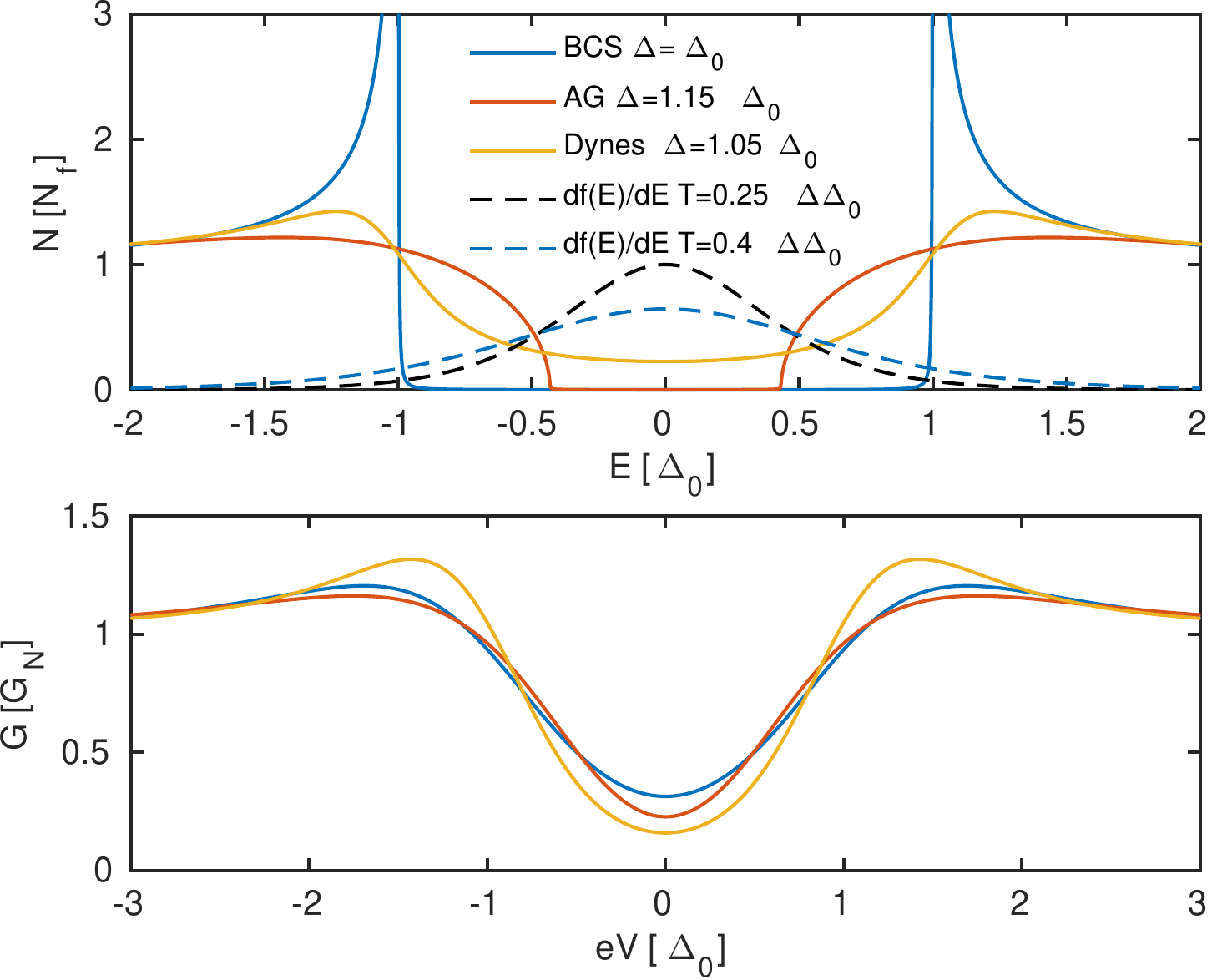}
	\caption{Top: The BCS density of states (blue, $\Delta=1$), the Abrikosov-Gor'kov one (red, $\Delta=1.15$, $\alpha=0.38$) and the Dynes one (yellow, $\Delta=1.05$, $\Gamma=0.24$). The dashed lines are the derivatives of the Fermi distribution function for $T=0.25\Delta$ (black) and $T=0.4\Delta$ (blue). Bottom: the corresponding $G(V)$ traces. The BCS DOS was convolved with the higher temperature distribution function, while the other two were convolved with the lower temperature one.
	}
	\label{fig:GV_toy_model}
\end{figure}

With the main goal of determining the order parameter as a function of the magnetic field, $\Delta(H)$, the following heuristic approach was adopted.
Without a proper microscopic theory to describe the soft gapped spectrum in presence of weak inter-valley scattering, and with limited ability to discern the details of the density of states (DOS) at $T=1.3\mathrm{K}$, the simplest approach to take is to try several different models for the $G(V)$ traces and compare the results. Even though there are countless models that one can utilize the discussion here is limited to the following three: an effective temperature $T^*$ with a BCS DOS, an Abrikosov-Gor'kov DOS (A-G DOS)  ~\cite{AG_DOS}  and a Dynes DOS   ~\cite{Dynes_DOS}. At finite temperatures the $G(V)$ curve is obtained by convolving the DOS with a distribution $\approx 3.5k_B T$ wide (FWHM): \large{$G(V) = \frac{1}{eR} \int_{-\infty}^{\infty} dE N(E) \frac{\partial f(E-V)}{\partial V}$}\normalsize, where $N(E)$ is the normalized density of states.  If $3.5k_B T \approx \Delta$ this leads to a finite conductance at zero voltage bias. The Abrikosov-Gor'kov depairing model lowers the spectral gap below the order parameter $\Delta$. When the depairing energy equals the order parameter, $\alpha(H)=\Delta(H)$ this leads to a gapless state, but even when $\alpha(H)<\Delta(H)$  along with a finite temperature the resulting $G(V)$ trace can be gapless. Lastly, the Dynes DOS is non-zero at $E=0$, directly leading to a soft gap. To illustrate this a simple example is shown in figure \ref{fig:GV_toy_model} - the top panel show the BCS, Abrikosov-Gor'kov and Dynes densities of states, as well as the distribution function $\frac{\partial f(E-V)}{\partial V}$ for some parameter values, while the bottom panel shows the corresponding differential conductance traces. Although the BCS and the Abrikosov-Gor'kov DOS' are fully gapped the resulting spectra are quite similar, and resemble a gapples spectrum.

\begin{figure}[h]
	\centering
	\includegraphics[width=0.7\textwidth]{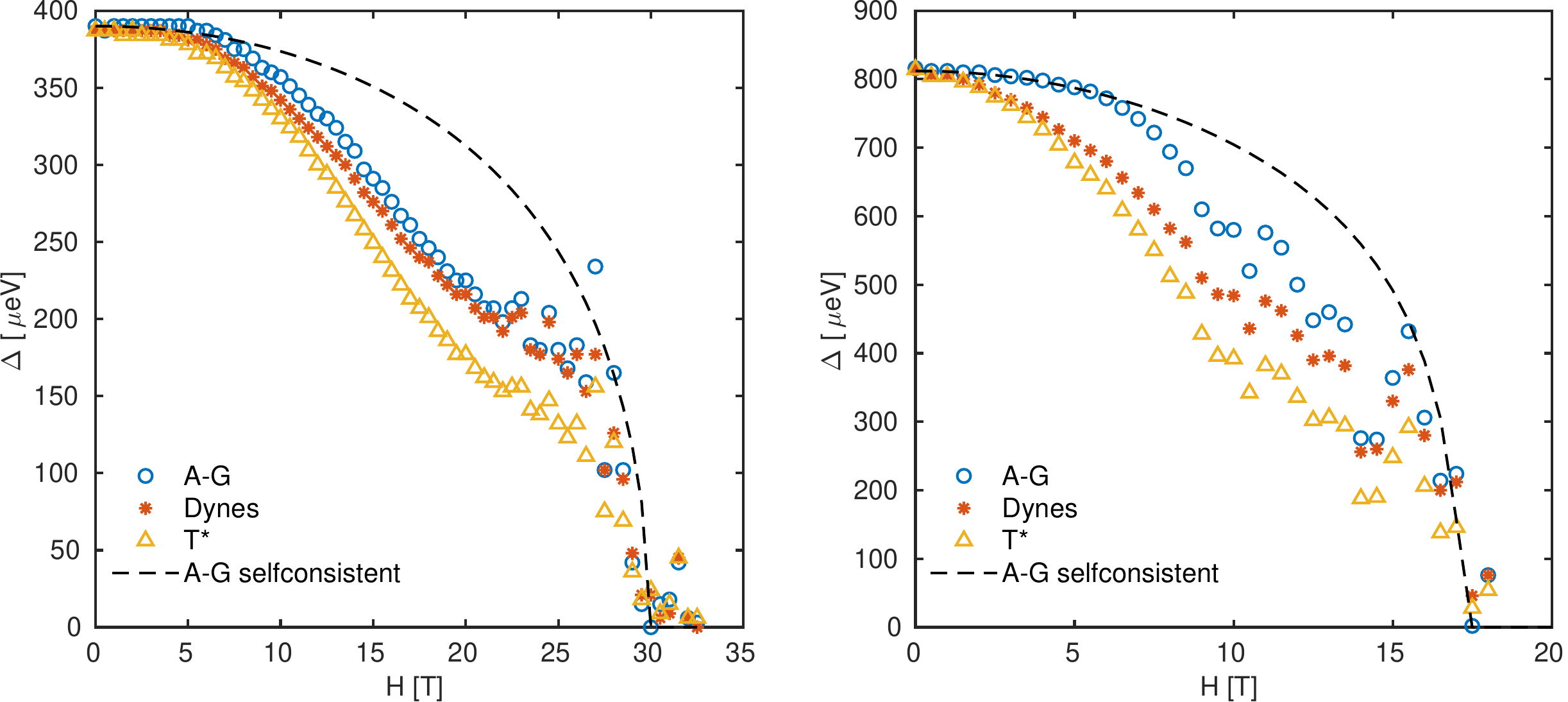}
	\caption{The extracted order parameters as function of the applied magnetic field using the three different $G(V)$ models, for the J7 (2ML, left panel), and J6 (4-8ML, right panel).
	}
	\label{fig:AG-Dynes-T}
\end{figure}

The parameters for the effective temperature model were the (field dependent) order parameter $\Delta(H)$ and the temperature $T^*(H)$. For the Abrikosov-Gor'kov and Dynes fits the temperature was fixed to $T=1.25\mathrm{K}$, while the depairing and Dynes energies were fitting parameters. The $\Delta(H)$ dependence obtained in this way, for both junctions and all three models, is shown on figure \ref{fig:AG-Dynes-T}.

\begin{figure}[h]
	\centering
	\includegraphics[width=0.5\textwidth]{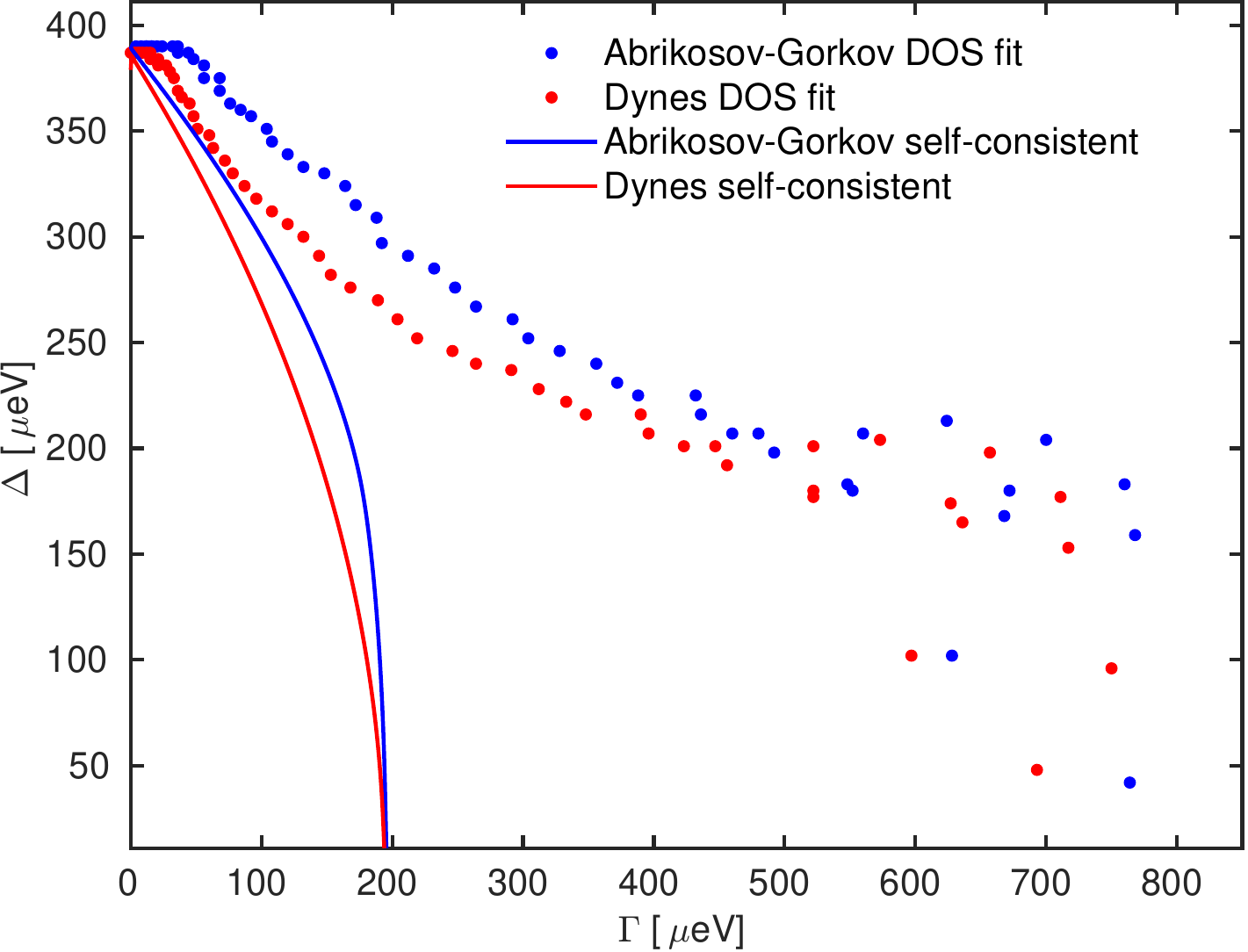}
	\caption{The order parameter $\Delta$ as a function of the Abrikosov-Gor'kov depairing (blue) and the Dynes energy (red) for J7, obtained from the fitting (dots) and the self-consistent gap equation (full lines).
	}
	\label{fig:AG_and_Dynes_SC_vs_fit}
\end{figure}

It is important to note that the values of these extra field dependent parameters, have no physical significance: the gap value is not self-consistently determined, nor should they be interpreted in the context of their usual meaning. They are rather just phenomenological parameters used to describe the obtained $G(V)$ spectra. To illustrate this figure \ref{fig:AG_and_Dynes_SC_vs_fit} shows the self-consistent and experimentally obtained $\Delta$ versus the Abrikosov-Gor'kov and Dynes $\Gamma$ parameters. 

\begin{figure}[h]
	\centering
	\includegraphics[width=0.5\textwidth]{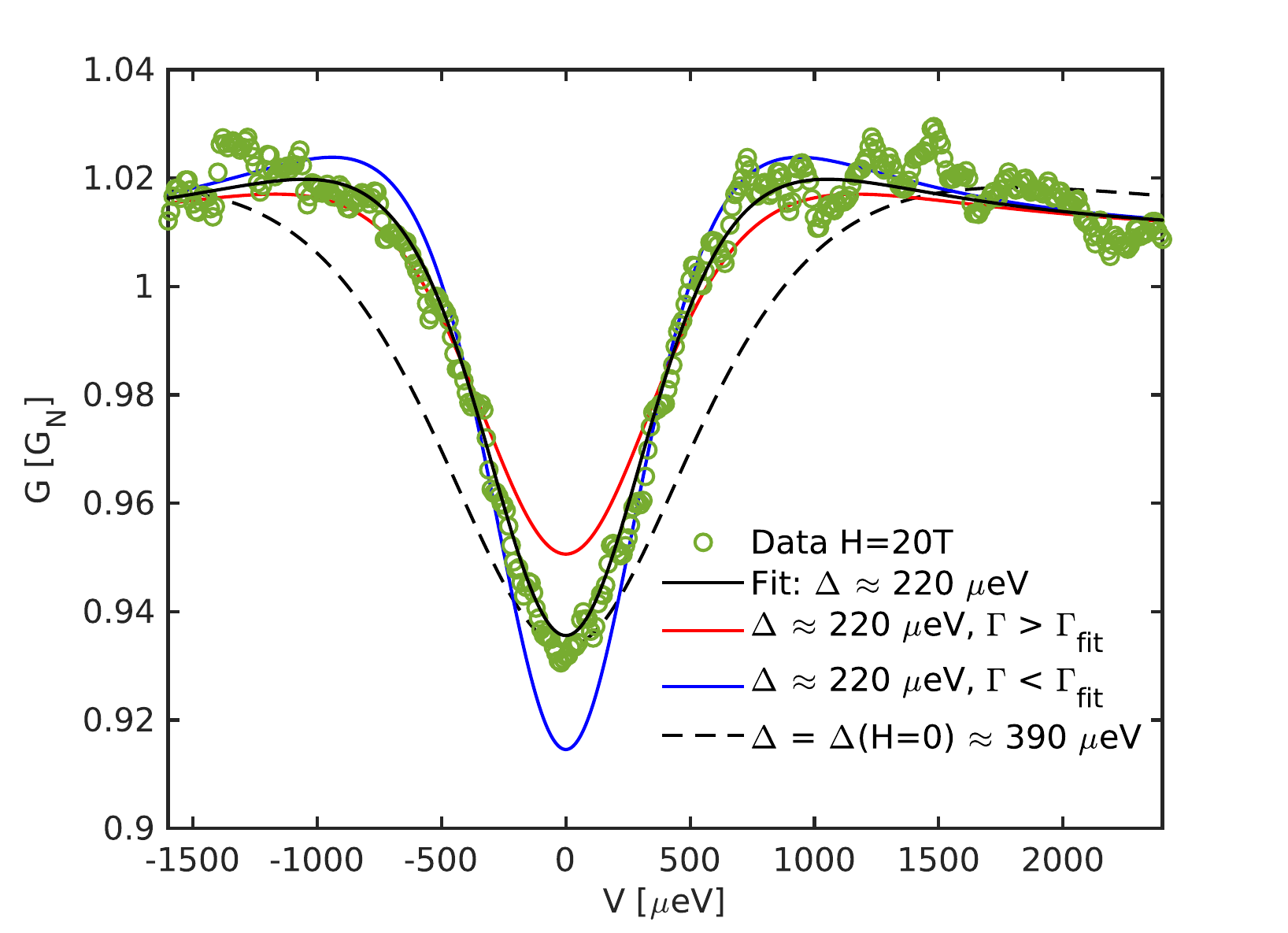}
	\caption{The experimental data from J7 at $H=20\mathrm{T}$ (green circles), the Abrikosov-Gor'kov fit (solid black) and two traces with the same $\Delta$ but different depairings and a trace with $\Delta=\Delta(H=0)$ with a depairing which fits the "depth" of the spectroscopic gap.
	}
	\label{fig:H20T_fit}
\end{figure}

As this approach estimates $\Delta$ not based on the details of the $G(V)$ spectrum, but rather the energy scale of the (spectroscopic) gap, it is important to show that this is a robust feature. To this end figure \ref{fig:H20T_fit} shows the $G(V)$ data from J7 at $H=20\mathrm{T}$, as well as several theoretical traces. The first of which is the Abrikosov-Gor'kov fit, followed by two traces with the same gap, but different depairing values, which demonstrate that the energy scale of the gap is dominantly set by $\Delta$ while the depth of the gap at $V=0$ is set by the depairing. The last trace shows that the gap is significantly different than the one found at $H=0$, countrary to what one might naively infer based on the colorplots shown in figure 3 of the main text.

Additionally the error of the $\Delta$ estimation can be performed in the following way: the log-likelihood distribution for the fitting parameters (given the data) is given by 
$$\tilde{p}(\Delta,\tilde{x}) = - \frac{ <(G_i - f(V_i,\Delta,\tilde{x}))^2>_i}{2\sigma^2}$$ where $\tilde{x}$ stands for the additional model-dependent parameters and $< ...>_i$ denotes the average over all of the acquired points. $\sigma$, the noise of the measurement, can be estimated either directly from  data, or  by the root-mean-square error (RMSE) of the fit, both of which give similar results. The log-likelihood is not necessarily quadratic in $\Delta$, as the fitting problem is nonlinear, but close to the maximum-likelihood point it can is approximately quadratic.
Therefore by fitting $\tilde{p}$ near it's maximal point with $-\frac{ (\Delta - \tilde{\Delta})^2}{2\delta_\Delta^2} + c$, where $c$ is related to the RMSE and $\tilde{\Delta}$ is the best fit value, we obtain an estimate of the fit uncertanty $\delta_\Delta$. The Abrikosov-Gor'kov $\Delta(H)$ curve with errors estimated in this way is shown in figure 4 of the main text. We find that the uncertainty of the extracted values of $\Delta$ is roughly equal to the spread of the data, regardless of the $G(V)$ model. The experimental $G(V)$ traces and the Abrikosov-Gor'kov fits are shown on figure 3 of the main text. The fits obtained using the other two models are almost indistinguishable from the Abrikosov-Gor'kov one. 

\subsection{DOS broadening}\label{partI:DOSbroadening}

Here we address the issue of in-gap states and the broadening of the coherence peaks. Indeed, the broadening parameters obtained from the fits are larger than expected from self-consistent Abrikosov-Gor'kov theory \cite{abrikosov1961zh} or a self-consistent Dynes model \cite{herman2016microscopic} (see figure \ref{fig:AG_and_Dynes_SC_vs_fit}). Moreover, at the critical field the broadening parameter is larger than theoretically allowed $2\alpha_c=\Delta(H=0)$. 

\begin{figure}[h!]
	\centering
	\includegraphics[width=0.5\textwidth]{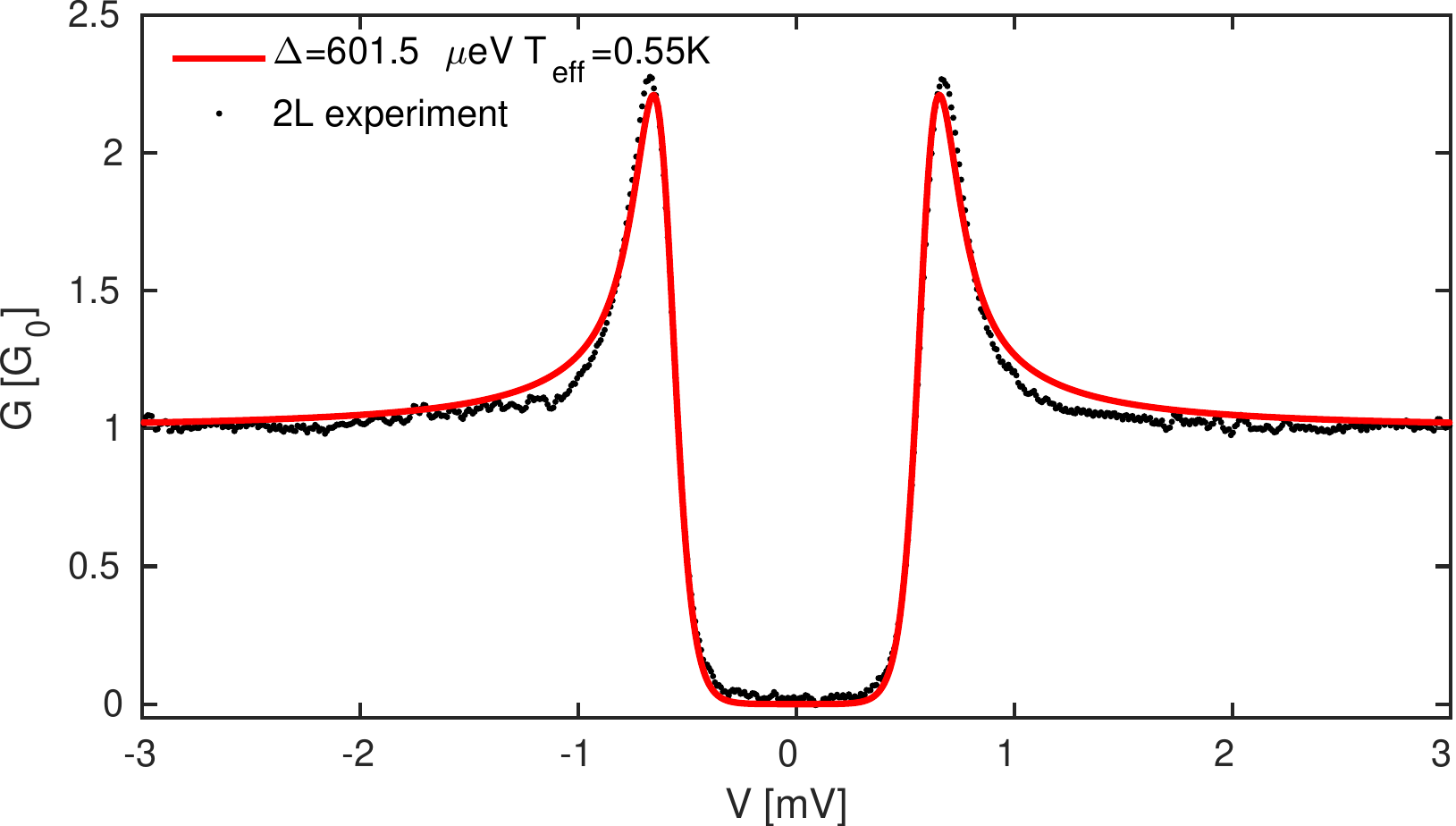}
	\caption{The bi-layer tunneling spectrum from \cite{khestanova2018unusual} (reproduced with the authors' permission), along with a BCS fit.}
	\label{fig:geim2L}
\end{figure}

There are several possible reasons for this unexpected DOS broadening: 
\begin{enumerate}
\item At the K/K' points, the two spins have slightly different dispersions and densities of states --- this can lead to broadening of the peak~\cite{marganska}. While Ref.~\cite{marganska} assumes superconductivity arising from Coulomb interactions, the same broadening phenomenon will appear for all pairing mechanisms. This is likely the reason why in both our work and that of \cite{khestanova2018unusual}, reproduced here in figure \ref{fig:geim2L}, the effective temperature is much higher than the measurement temperature. This is the case even at (higher) temperatures where there should be good electron-phonon coupling, and where the electron temperature should thus be that of the refrigerator ($\sim$ 100 mK and above). 

Besides this, the following could also contribute to the broadening of the DOS.
\item In-gap states may appear in triplet superconductors because of field-induced gap nodes at the Fermi surface~\cite{fischer2018}. Such nodes would occur along the $\Gamma$-M directions when $E_Z$ (the Zeeman energy) exceeds the order parameter $\Delta_s$. This means that only the $\Gamma$ pockets would be affected and not the K/K' ones. Since electrons tunnel into the K/K' pockets (see Section IIC1), some K/K'-$\Gamma$ coupling would have to be present for such nodes to affect our measured density of states. 
\item Inelastic tunneling; energy loss due to defects in the barrier will broaden the $G(V)$ characteristic~\cite{hlobil,reed}. 
\item Coupling of the charge density waves via phonons either to the quasiparticles, or to amplitude fluctuations (Higgs mode) of the order parameters~\cite{littlewood1982}. 
\end{enumerate}

These various factors are difficult to disentangle; however, none of them significantly affects the energy gap, which can be determined with high precision from the fits, as was shown in this section.

\FloatBarrier

\subsection{Estimate of Inter-valley (K/K') and Total Scattering Times}\label{partI:scattering}

\textit{Intervalley Scattering Time:} Ref.~\cite{ilic2017} gives $H_{||}^c$ as a function of temperature for different $\tau_{iv}$, the scattering time between K and K' points, for an Ising superconductor. While the theory in Ref.~\cite{ilic2017} does not include a triplet order parameter or two-pocket coupling, this should not change the order of magnitude of estimates of $H_{||}^c$ for a given $\tau_{iv}$ and vice versa. (This can be seen in Figure 1c of the main text.) In our case $E_{SO}/(k_B T_c) \sim 20$. From Figure 5c in Ref.~\cite{ilic2017}, and using our experimental values for $T$, $T_c$ and $H_{||}^c$ we estimate $h/\tau_{iv} \approx k_B T_{cs}$. $h/\tau_{iv}$ is therefore on the order of $\Delta_s \approx 400 \mu$eV. 

\textit{Total Scattering Time:} In Ref.~\cite{barrera2018}, the mean free path $\ell$ for a bilayer NbSe$_2$ device was estimated to be 17 nm from Hall measurements. Thus, the total scattering time, including all intra- and inter-valley processes, is $\tau = \ell/v_F$. As $v_F \approx 5 \cdot 10^4$ m/s~\cite{dvir2018spectroscopy}, this gives $\tau \approx 340$ fs. Now, our bilayer device is more disordered than that of Ref.~\cite{barrera2018}: whereas $H_c^{||}$ in Ref.~\cite{barrera2018} is the same as in our case, their critical temperature is 5K, as opposed to 2.6K for us. Thus, $1/\tau \approx 10$ meV can be considered a lower limit for the total scattering time $1/\tau$ in our bilayer device. This is consistent with the estimate above of the inter-valley scattering rate, that is to say $1/\tau > 1/\tau_{iv}$. 

From the above two estimates, and from the fact that single-particle $\Gamma-K/K'$ scattering (McMillan coupling) does not seem to be important for our devices (cf. Section IID below), we conclude that intra-pocket (K-K, K'-K' or $\Gamma-\Gamma$) is the dominant disorder-induced scattering.

\section{Theory}\label{partII}

In this appendix, we  calculate the superconducting
energy gap of a monolayer superconductor without an inversion
center and in the presence of an in-plane magnetic field.  
In Sec. II.A, we introduce the appropriate Hamiltonian. In Sec. II. B, we study the density of states in the presence of singlet and triplet pairing in a simplified model restricted to the K/K’ pockets. In Secs. II.C \& II.D, we address the interplay between the K/K’ and $\Gamma$ pockets within two different models taking into account singlet pairing only : in Sec. II.C, we use the so-called Suhl-Matthias-Walker coupling describing inter-pocket pairing or Cooper pair tunneling, whereas in Sec. II.D, we use the so-called MacMillan coupling describing inter-pocket single-particle tunneling or scattering.


\subsection{The model Hamiltonian}\label{partII:hamiltonian}
\label{sec:model} 

The model Hamiltonian is a sum of the free part $H_0$ which defines our band structure model, the disorder Hamiltonian $H_{dis}$ and the interaction $H_{\mathrm{int}}$, 
\begin{equation}
H=H_{0}+ H_{dis} + H_{\mathrm{int}}. \label{eq:8.1}
\end{equation}
Our band structure model includes the Nb derived bands.
The crossing of the Nb derived band with the Fermi level gives rise to a hole-like pocket centered at $\Gamma$ as well as a a pair of hole-like pockets centered at $K(K')$ labeled by the valley index, $\eta=\pm 1$. 
In result, we have
\begin{equation}
H_{0}=\sum_{\eta}\underset{\mathbf{k},s}{\sum}\xi_{\mathbf{k}}^{K}a_{\eta\mathbf{k}s}^{\dagger}a_{\eta\mathbf{k}s}+\sum_{\eta}\underset{\mathbf{k},s,s'}{\sum}\left(\boldsymbol{\gamma}_{\eta}^{K}-\mathbf{B}\right)\cdot\boldsymbol{\sigma}_{ss'}a_{\eta\mathbf{k}s}^{\dagger}a_{\eta\mathbf{k}s'}+\underset{\mathbf{k},s}{\sum}\xi_{\mathbf{k}}^{\Gamma}b_{\mathbf{k}s}^{\dagger}b_{\mathbf{k}s}+\underset{\mathbf{k},s,s'}{\sum}\left(\boldsymbol{\gamma}_{\mathbf{k}}^{\Gamma}-\mathbf{B}\right)\cdot\boldsymbol{\sigma}_{ss'}b_{\mathbf{k}s}^{\dagger}b_{\mathbf{k}s'}.\label{eq:8.2}
\end{equation}
Here $a_{\eta\mathbf{k}s}^{\dagger}$ and $b_{\mathbf{k}s}^{\dagger}$
are the creation operators in the $K/K'$ and $\Gamma$ pockets, respectively,
$\xi_{\mathbf{k}}^{\Gamma\left(K\right)}$ is the spin-independent part of the band energy measured relative to Fermi energy,
$\boldsymbol{\sigma}=\left(\sigma_{x},\sigma_{y},\sigma_{z}\right)$
is the vector of Pauli matrices, $\boldsymbol{\gamma}_{\bar{\eta}}^{K}=-\boldsymbol{\gamma}_{\eta}^{K}$ and $\boldsymbol{\gamma}_{\bar{\mathbf{k}}}^{\Gamma}=-\boldsymbol{\gamma}_{\mathbf{k}}^{\Gamma}$ 
are the anti-symmetric SOC terms arising from the lack of inversion
symmetry and $\mathbf{B}=E_{Z}\hat{x}$ is the Zeeman field, i.e. the magnetic field which
absorbs the prefactor $g\mu_{\mathrm{B}}/2$ that includes the $g$-factor
and Bohr magneton. 
We use the notation $\bar{\eta}=-\eta,\bar{\mathbf{k}}=-\mathbf{k}$.
For the $K$-pocket and $\Gamma$-pocket we consider the Ising SOC of the form,
\begin{align}\label{eq:SOC_KG}
\boldsymbol{\gamma}_{\eta}^{K}=\eta E_{\mathrm{SO}}^{K}\hat{z},\ \ \ \boldsymbol{\gamma}_{\mathbf{k}}^{\Gamma}=E_{\mathrm{SO}}^{\Gamma}\cos\left(3\varphi_{\hat{\mathbf{k}}}\right)\hat{z},
\end{align}
where $\varphi_{\hat{\mathbf{k}}}$ is the angle formed by the momentum unit vector ${\hat{\mathbf{k}}}$ of an electron with the $k_x$ direction. 
Within the $\Gamma$ pocket, the SOC $\boldsymbol{\gamma}_{\mathbf{k}}^{\Gamma}$, Eq.~\eqref{eq:SOC_KG}, changes sign six times at the $\Gamma M$ lines.
In the $K$ and $K'$ pockets, it is constant and antiparallel.


The presence of randomly
distributed scalar impurities gives rise to a scattering potential,
\begin{align}
H_{dis}(\mathbf{r}) = \sum_{l} U_0(\mathbf{r} - \mathbf{R}_l)    \, ,   
\end{align}
where $\mathbf{R}_l$ is the location of the $l$th impurity scattering center. 
Singlet pairing as well as inter-valley triplet pairing in K/K' pockets is not sensitive to intra-valley scattering. 
To study the effect of inter-valley scattering, we consider a short-range impurity potential such that $U_{0}\left(\mathbf{k}-\mathbf{k}'\right)=U_{0}$. The effect of the impurity potential is described within the self-consistent Born approximation by the appropriate self-energy $\hat{\Sigma}$, which we do not write here explicitly. $\hat{\Sigma}$ is proportional to the scattering rate $1/\tau_{iv}=\pi n_{imp}N_{0}U_{0}^{2}$, where $n_{imp}$ is the impurity density and  $N_{0}$ is the normal state density of states per spin species. Superconductivity within the the $\Gamma$ pocket is sensitive to intra-pocket scattering at finite magnetic fields. We characterize it by a  scattering rate $1/\tau_\Gamma$. As it involves a small momentum transfer, both short-and long-range impurities contribute. Thus we expect $1/\tau_\Gamma\gg1/\tau_{iv}$. In the following, we set $k_{\mathrm{B}}=1$.

The interaction Hamiltonian, $H_{\mathrm{int}}$ in Eq.~\eqref{eq:8.1}, contains the superconducting pairing interactions
\begin{align}
\label{eq:Hint}
H_{\mathrm{int}} & =\frac{1}{2}\sum_{\eta,\eta'}\underset{s_{i},s_{i}'}{\sum}\underset{\mathbf{k},\mathbf{k}'}{\sum}V_{s_{1}s_{2},s_{1}'s_{2}'}^{KK}\left(\mathbf{k},\eta;\mathbf{k}',\eta'\right)a_{\eta\mathbf{k}s_{1}}^{\dagger}a_{\bar{\eta}\bar{\mathbf{k}}s_{2}}^{\dagger}a_{\bar{\eta}'\bar{\mathbf{k}}'s_{2}'}a_{\eta'\mathbf{k}'s_{1}'}+\frac{1}{2}\underset{s_{i},s_{i}'}{\sum}\underset{\mathbf{k},\mathbf{k}'}{\sum}V_{s_{1}s_{2},s_{1}'s_{2}'}^{\Gamma\Gamma}\left(\mathbf{k},\mathbf{k}'\right)b_{\mathbf{k}s_{1}}^{\dagger}b_{\bar{\mathbf{k}}s_{2}}^{\dagger}b_{\bar{\mathbf{k}}'s_{2}'}b_{\mathbf{k}'s_{1}'}\\
 & +\frac{1}{2}\sum_{\eta}\underset{s_{i},s_{i}'}{\sum}\underset{\mathbf{k},\mathbf{k}'}{\sum}V_{s_{1}s_{2},s_{1}'s_{2}'}^{K\Gamma}\left(\mathbf{k},\eta;\mathbf{k}'\right)a_{\eta\mathbf{k}s_{1}}^{\dagger}a_{\bar{\eta}\bar{\mathbf{k}}s_{2}}^{\dagger}b_{\bar{\mathbf{k}}'s_{2}'}b_{\mathbf{k}'s_{1}'}+\frac{1}{2}\sum_{\eta'}\underset{s_{i},s_{i}'}{\sum}\underset{\mathbf{k},\mathbf{k}'}{\sum}V_{s_{1}s_{2},s_{1}'s_{2}'}^{\Gamma K}\left(\mathbf{k};\mathbf{k}',\eta'\right)b_{\mathbf{k}s_{1}}^{\dagger}b_{\bar{\mathbf{k}}s_{2}}^{\dagger}a_{\bar{\eta}'\bar{\mathbf{k}}'s_{2}'}a_{\eta'\mathbf{k}'s_{1}'}.\nonumber 
\end{align}
where $V^{KK}$ and $V^{\Gamma\Gamma}$ are the intra-pocket
pairing interactions in the $K(K')$ and $\Gamma$ pockets, respectively, whereas $V^{K\Gamma}$ $V^{\Gamma K}$ are the inter-pocket pairing interactions.

We introduce the two order parameters (OPs) in the standard way, 
\begin{align}
\label{eq:OP}
\underline{\Delta}_{s_{1}s_{2}}^{K,\eta}\left(\mathbf{k}\right)=\frac{1}{V}\underset{\mathbf{k}',s_{1}',s_{2}'}{\sum}\left[\sum_{\eta'}V_{s_{1}s_{2},s_{1}'s_{2}'}^{KK}\left(\mathbf{k},\eta;\mathbf{k}',\eta'\right)\left\langle a_{\bar{\eta}'\bar{\mathbf{k}}'s_{2}'}a_{\eta'\mathbf{k}'s_{1}'}\right\rangle +V_{s_{1}s_{2},s_{1}'s_{2}'}^{K\Gamma}\left(\mathbf{k},\eta;\mathbf{k}'\right)\left\langle b_{\bar{\mathbf{k}}'s_{2}'}b_{\mathbf{k}'s_{1}'}\right\rangle \right],
\notag \\
\underline{\Delta}_{s_{1}s_{2}}^{\Gamma}\left(\mathbf{k}\right)=\frac{1}{V}\underset{\mathbf{k}',s_{1}',s_{2}'}{\sum}\left[V_{s_{1}s_{2},s_{1}'s_{2}'}^{\Gamma\Gamma}\left(\mathbf{k};\mathbf{k}'\right)\left\langle b_{\bar{\mathbf{k}}'s_{2}'}b_{\mathbf{k}'s_{1}'}\right\rangle +\sum_{\eta'}V_{s_{1}s_{2},s_{1}'s_{2}'}^{\Gamma K}\left(\mathbf{k};\mathbf{k}',\eta'\right)\left\langle a_{\bar{\eta}',\bar{\mathbf{k}}',s_{2}'}a_{\eta',\mathbf{k}',s_{1}'}\right\rangle \right], 
\end{align}
where $V$ is the volume of the system, and $\langle \ldots \rangle$ stands for the thermodynamic and quantum mechanical averaging.
They can be represented in the standard matrix form,
\begin{equation}
\underline{\Delta}^{\Gamma(K,\eta)}\left(\mathbf{k}\right)
=\left[\psi^{\Gamma(K,\eta)}\left(\mathbf{k}\right)+\mathbf{d}^{\Gamma(K,\eta)}\left(\mathbf{k}\right)\cdot\boldsymbol{\sigma}\right]i\sigma_{2}.\label{eq:1.8}
\end{equation}
Here $^{\Gamma(K,\eta)}\left(\mathbf{k}\right)$ and $\mathbf{d}^{\Gamma(K,\eta)}\left(\mathbf{k}\right)$
parametrize the singlet and triplet components of the OP, respectively.
For simplicity, we assume the singlet OPs to be isotropic,
$\psi^{\Gamma\left(K,\eta\right)}\left(\mathbf{k}\right)=\Delta_{s}^{\Gamma\left(K\right)}$.

Interactions in Eq.~\eqref{eq:Hint} may have components in both singlet and triplet channels, 
\begin{align}
    V^{\beta\beta'}= V_{s}^{\beta\beta'} + V_{t}^{\beta\beta'}
\end{align}
where $\beta,\beta'$ refer to the pocket index. For the singlet pairing we take 
\begin{align}\label{eq:Vsinglet}
V_{s}^{\beta\beta'}=v^{\beta\beta'}\left[i\sigma_{y}\right]_{s_{1}s_{2}}\left[i\sigma_{y}\right]_{s_{1}'s_{2}'}^{*}.
\end{align}
The hermiticity condition yields $\left[v^{\Gamma \Gamma}\right]^* = v^{\Gamma \Gamma}$, $\left[v^{K K}\right]^* = v^{K K} $, and $\left[v^{\Gamma K}\right]^* = v^{K \Gamma}$.
Furthermore, we take inter-pocket couplings real which makes them equal,
$v^{\Gamma K} = v^{K \Gamma}$, and consider attractive interactions, $v^{\beta\beta'}< 0$ for definitness.

In this SI, the triplet OP will be considered in the $K/K'$ pockets only, $\mathbf{d}^{K} = \mathbf{d}_\eta$,
\begin{equation}
\mathbf{d}_\eta=\hat{\gamma}_{\eta}\left(\eta_{E1}\hat{x}+\eta_{E2}\hat{y}+\eta_{A}\hat{z}\right)\label{eq:1.9}
\end{equation}
where $\eta_{A}$ and $\eta_{E1(2)}$ are triplet OPs transforming trivially and non-trivially under $D_{3h}$.
Eq.~\eqref{eq:1.9} leads us to the effective interaction,
\begin{align}\label{eq:1.10}
V^{KK}_t(\eta,\eta') & =
\underset{j=1,2}{\sum}v_{t}\left[\hat{\gamma}_{\eta}\sigma_{j}i\sigma_{2}\right]_{s_{1}s_{2}}\left[\hat{\gamma}_{\eta'}\sigma_{j}i\sigma_{2}\right]_{s_{1}'s_{2}'}^{*}
+v_{tz}\left[\hat{\gamma}_{\eta}\sigma_{3}i\sigma_{2}\right]_{s_{1}s_{2}}\left[\hat{\gamma}_{\eta'}\sigma_{3}i\sigma_{2}\right]_{s_{1}'s_{2}'}^{*}.
\end{align}

As shown in Ref.  \cite{Mockli2019}, the magnetic field couples the singlet order parameter $\Delta_s$ and the equal spin triplet order parameter $\eta_{E2}$.
For the purpose of fitting the data, we assume that singlet pairing is dominant and that the temperature is larger than the critical temperature of all the possible triplet pairings. In that case, we can set $\eta_A=\eta_{E1}=0$ and keep only the singlet and the $\eta_{E2}$-triplet order parameters.

 We define the transition temperature $T_{cs}$ ($T_{ct}$)
by setting $E_Z=E_{\mathrm{SO}}=\Gamma=0$ and keeping only the
$\Delta_s$($\eta_{E2}$) OP in Eq. \eqref{eq:1.8}.
The relation between $T_{cs}$ and $v_{s}$ is $T_{cs}=2\Lambda e^{\gamma_{E}}\pi^{-1}\exp\left[-1/2N\left|v_{s}\right|\right]$,
where $\Lambda$ is a cutoff for the high energy attraction, $N$ is the density of states per spin summed over all superconducting pockets, and $\gamma_{E}$
is Euler's constant. Similarly, for $T_{ct}$, we have $T_{ct}=2\Lambda e^{\gamma_{E}}\pi^{-1}\exp\left[-1/2N\left|v_{t}\right|\right]$.
For the analysis, we use the transition temperatures rather than the interaction
amplitudes as parameters.

The Bogoliubov-de Gennes (BdG) Hamiltonian for electrons in the $K/K'$ and $\Gamma$ pockets reads
\begin{equation}
\label{eq:BdG}
\hat{H}_{\mathrm{BdG}}^{\Gamma\left(K,\eta\right)}=\left[\begin{array}{cc}
\xi_{\mathbf{k}}+\left[\boldsymbol{\gamma}_{\mathbf{k}\left(\eta\right)}^{\Gamma\left(K\right)}-\mathbf{B}\right]\cdot\boldsymbol{\sigma} & \underline{\Delta}^{K\left(\Gamma\right)}\\
\underline{\Delta}^{\Gamma\left(K\right)\dagger} & -\xi_{\mathbf{k}}+\left[\boldsymbol{\gamma}_{\mathbf{k}\left(\eta\right)}^{\Gamma\left(K\right)}+\mathbf{B}\right]\cdot\boldsymbol{\sigma}^{\mathrm{T}}
\end{array}\right]\, .
\end{equation}
Now, we will consider different scenarios: In subsection \ref{sec-KK'}, we will consider both singlet and triplet pairing in a simplified model neglecting the $\Gamma$ pocket.
In subsections \ref{sec:two-pockets-MSW} and \ref{sec-mm}, we consider the effect of the $\Gamma$ pocket, but neglecting triplet pairing.

\subsection{Simplified model without the $\Gamma$ pocket}
\label{sec-KK'}

Focusing on the $K$ pocket with triplet component of the OP, the dispersion relation is determined by the solution of the equation
\begin{equation}
\mathrm{det}\left[E\hat{\sigma}_{0}-\hat{H}_{\mathrm{BdG}}^{K,\eta}\right]=0\label{eq:1.20}
\end{equation}
 for $E$, where $\hat{\sigma}_{0}$ is the $4\times4$ unit matrix. We choose
the phase of the singlet OP $\Delta_s$ to be $0$. The coupling between the singlet and triplet order parameters imposes their relative phases such that $\mathbf{d}_\eta=i\,{\rm sign}(E_Z)\hat{\gamma}_\eta\Delta_{tB}\hat{y}$ with $\Delta_{tB}$ real and positive. The physically
relevant solution for the energy is
\begin{align}
E\left(\xi_{\mathbf{k}}\right)  =&\biggr(\xi_{\mathbf{k}}^{2}+E_{\mathrm{SO}}^{2}+E_Z^{2}+\Delta_s^2+\Delta_{tB}^{2}\label{eq:1.21}\\
 & -2\sqrt{\xi_{\mathbf{k}}^{2}\left(E_{\mathrm{SO}}^2+E_Z^{2}\right)+\left(|E_Z|\Delta_s-E_{\mathrm{SO}}\Delta_{tB}\right)^{2}}\biggr)^{1/2}.\nonumber 
\end{align}
The dispersion $E\left(\xi_{\mathbf{k}}\right)$ has a minimum at
$\xi_{\mathbf{k}}=\sqrt{\rho^2-P^{2}/\rho^{2}}$, where we introduced the notation $\rho=\sqrt{E_{\mathrm{SO}}^{2}+E_Z^2}$ and  $P=|E_Z|\Delta_s-E_{\mathrm{SO}}\Delta_{tB}$, which gives 
the superconducting energy gap
\begin{eqnarray}
\Delta&=&\frac1{\rho}\left(E_{\mathrm{SO}}\Delta_s+|E_Z|\Delta_{tB}\right).\label{eq:1.22}
\end{eqnarray}

\subsubsection{Without inter-valley scattering: Quasiclassical Green functions}

The coupled order parameters and  density of states can be computed using quasiclassical Green functions. In the absence of inter-valley scattering, we obtain~\cite{Ilic-unpub}
\begin{eqnarray}
\nu(E)
&=&2N_0\Re\left[\frac{\omega_n{\rm\, sign}(\Sigma)}{\sqrt2\left[\Sigma-2\rho^2 +{\rm\, sign}(\Sigma)\sqrt{\Sigma^2-4P^2}\right]^{1/2}}\left(1+\frac{|\Sigma|}{\sqrt{\Sigma^2-4P^2}}\right)\right]_{i\omega_n\to E+i\delta},\label{eq-dosf}\\
\Delta_s
&=&2\pi T|v_s|\sum_{\omega_n>0} \frac1{\sqrt2\left[\Sigma-2\rho^2 +\sqrt{\Sigma^2-4P^2}\right]^{1/2}}\left[\Delta_s\left(1+\frac{\Sigma-2E_Z^2}{\sqrt{\Sigma^2-4P^2}}\right)+\Delta_{tB}\frac{2|E_Z|E_{\mathrm{SO}}}{\sqrt{\Sigma^2-4P^2}}\right],\label{eq-s}\\
\Delta_{tB}
&=&2\pi T|v_t|\sum_{\omega_n>0} \frac1{\sqrt2\left[\Sigma-2\rho^2 +\sqrt{\Sigma^2-4P^2}\right]^{1/2}}\left[\Delta_{tB}\left(1+\frac{\Sigma-2E_{\mathrm{SO}}^2}{\sqrt{\Sigma^2-4P^2}}\right)+\Delta_s\frac{2|E_Z|E_{\mathrm{SO}}}{\sqrt{\Sigma^2-4P^2}}\right],\label{eq-t}
\end{eqnarray}
where $\omega_{n}=\pi T\left(2n+1\right)$ are fermionic Matsubara frequencies and we introduced the notation $\Sigma=\omega_n^2+\rho^2+\Delta_s^2+\Delta_{tB}^2$.

The density of states Eq.~\eqref{eq-dosf} displays a superconducting energy gap $\Delta$. Furthermore, there is  a partial \lq\lq mirage\rq\rq\ gap~\cite{Belzig}  centered around $E=\pm\sqrt{\rho^2+\Delta_s^2+\Delta_{tB}^2}$.

In general, the coupled self-consistency equations can be solved numerically. The fit shown in the main text was obtained that way. However, simplifications are possible in the limit $E_{\mathrm{SO}}\gg\Delta_0$, where $\Delta_0$ is the zero-temperature, zero-field singlet order parameter. In that case, Eqs.~\eqref{eq-s} and \eqref{eq-t} may be combined into one equation for the gap $\Delta$,
\begin{eqnarray}
\!\!\frac{\left[2\pi T|v_s|\sum_{\omega_n>0} \frac1{\sqrt{\omega_n^2+\Delta^2}}-1\right]\left[2\pi T|v_t|\sum_{\omega_n>0} \frac1{\sqrt{\omega_n^2+\Delta^2}}-1\right]}{|v_t|E_{\mathrm{SO}}^2\left[2\pi T|v_s|\sum_{\omega_n>0} \frac1{\sqrt{\omega_n^2+\Delta^2}}-1\right]+|v_s|E_Z^2\left[2\pi T|v_t|\sum_{\omega_n>0} \frac1{\sqrt{\omega_n^2+\Delta^2}}-1\right]}&=&2\pi T\sum_{\omega_n>0}\frac1{\sqrt{\omega_n^2+\Delta^2}}\frac1{\omega_n^2+\rho^2}.\;\;\label{eq-gap}
\end{eqnarray}
The density of states  at $|E|\ll E_{\mathrm{SO}}$ acquires a BCS form,
\begin{equation}
\nu(E)=\nu_0\frac{|E|}{\sqrt{E^2-\Delta^2}}\theta(|E|-\Delta).
\end{equation}

\subsubsection{With inter-valley scattering: Landau expansion}

To study the effect of disorder on a qualitative level, we employ a Landau expansion valid close to the critical temperature. 
Considering the model Hamiltonian above and using quasiclassical methods
the difference of the thermodynamic potential in superconducting and
normal state $\Omega$ may be written in the form of a Landau expansion
as
\begin{equation}
\left(V^{2}N_{0}\right)^{-1}\Omega\left(\Delta_s,\Delta_{tB}\right)=\Omega^{\left(2\right)}+\Omega^{\left(4\right)}.\label{eq:1.11}
\end{equation}
Here $\Omega^{\left(2\right)}$ contains the terms quadratic in the
OPs, and $\Omega^{\left(4\right)}$ contains the quartic terms. For $\Omega^{\left(2\right)}$, we have
\begin{align}
\Omega^{\left(2\right)} & =A_{1}\Delta_s^{2}+A_{2}\Delta_{tB}^{2}+2A_{3}\Delta_s\Delta_{tB}. \label{eq:1.12}
\end{align}

Denoting $\tilde{\omega}_{n}=\omega_{n}+\mathrm{sgn}\left(\omega_{n}\right)1/\tau_{iv}$,
the coefficients are given as~\cite{Mockli2020}, 
\begin{equation}
A_{1}=2\pi T\underset{\omega_{n}>0}{\sum}\frac{\tilde{\omega}_{n}E_Z^{2}}{\omega_{n}\left[\tilde{\omega}_{n}\left(E_Z^{2}+\omega_{n}^{2}\right)+\omega_{n}E_{\mathrm{SO}}^{2}\right]}+\ln\frac{T}{T_{cs}},\label{eq:1.13}
\end{equation}
\begin{equation}
A_{2}=2\pi T\underset{\omega_{n}>0}{\sum}\frac{\tau_{iv}^{-1}\left(E_Z^{2}+\omega_{n}^{2}\right)+\omega_{n}E_{\mathrm{SO}}^{2}}{\omega_{n}\left[\tilde{\omega}_{n}\left(E_Z^{2}+\omega_{n}^{2}\right)+\omega_{n}E_{\mathrm{SO}}^{2}\right]}+\ln\frac{T}{T_{ct}},\label{eq:1.14}
\end{equation}
\begin{equation}
A_{3}=2\pi T\underset{\omega_{n}>0}{\sum}\frac{(-E_Z) E_{\mathrm{SO}}}{\tilde{\omega}_{n}\left(E_Z^{2}+\omega_{n}^{2}\right)+\omega_{n}E_{\mathrm{SO}}^{2}}.\label{eq:1.16}
\end{equation}

For succinctness, we do not provide here the full expression for $\Omega^{\left(4\right)}$
and only write the term corresponding to the singlet OP 
\begin{equation}
\Omega_{s}^{\left(4\right)}=-\pi T_{cs}\underset{\omega_{n}'>0}{\sum}D_{1}\left(\omega_{n}'\right)\Delta_s^4,\label{eq:1.17}
\end{equation}
where
\begin{align}
D_{1}\left(\omega\right) & =\frac{1}{2\left[\tilde{\omega}\left(E_Z^{2}+\omega^{2}\right)+\omega E_{\mathrm{SO}}^{2}\right]^{4}}\biggr[-\omega\left(\omega\tilde{\omega}+E_{\mathrm{SO}}^{2}\right)^{4}\label{eq:1.18}\\
 & +2E_Z^{2}\omega\left(\tilde{\omega}^{2}-E_{\mathrm{SO}}^{2}\right)\left(\omega\tilde{\omega}+E_{\mathrm{SO}}^{2}\right)^{2}\nonumber \\
 & +E_Z^{4}\left(3\omega\tilde{\omega}^{4}+2E_{\mathrm{SO}}^{2}\tilde{\omega}^{2}\left(\tau_{iv}^{-1}+\tilde{\omega}\right)-\omega E_{\mathrm{SO}}^{4}\right)\biggr].\nonumber 
\end{align}
and where $\omega_{n}'=\pi T_{cs}\left(2n+1\right)$. In the limit $E_{\mathrm{SO}}=0$ and no triplet OP, $\Delta_{tB}=0$, derivation of the thermodynamic potential \eqref{eq:1.11} by the singlet OP reproduces the self consistency equation found in Ref.~\cite{Maki1964}. Equipped
with the thermodynamic potential, the
OPs are found by the process of minimization. 

\begin{figure}
\includegraphics[scale=0.51]{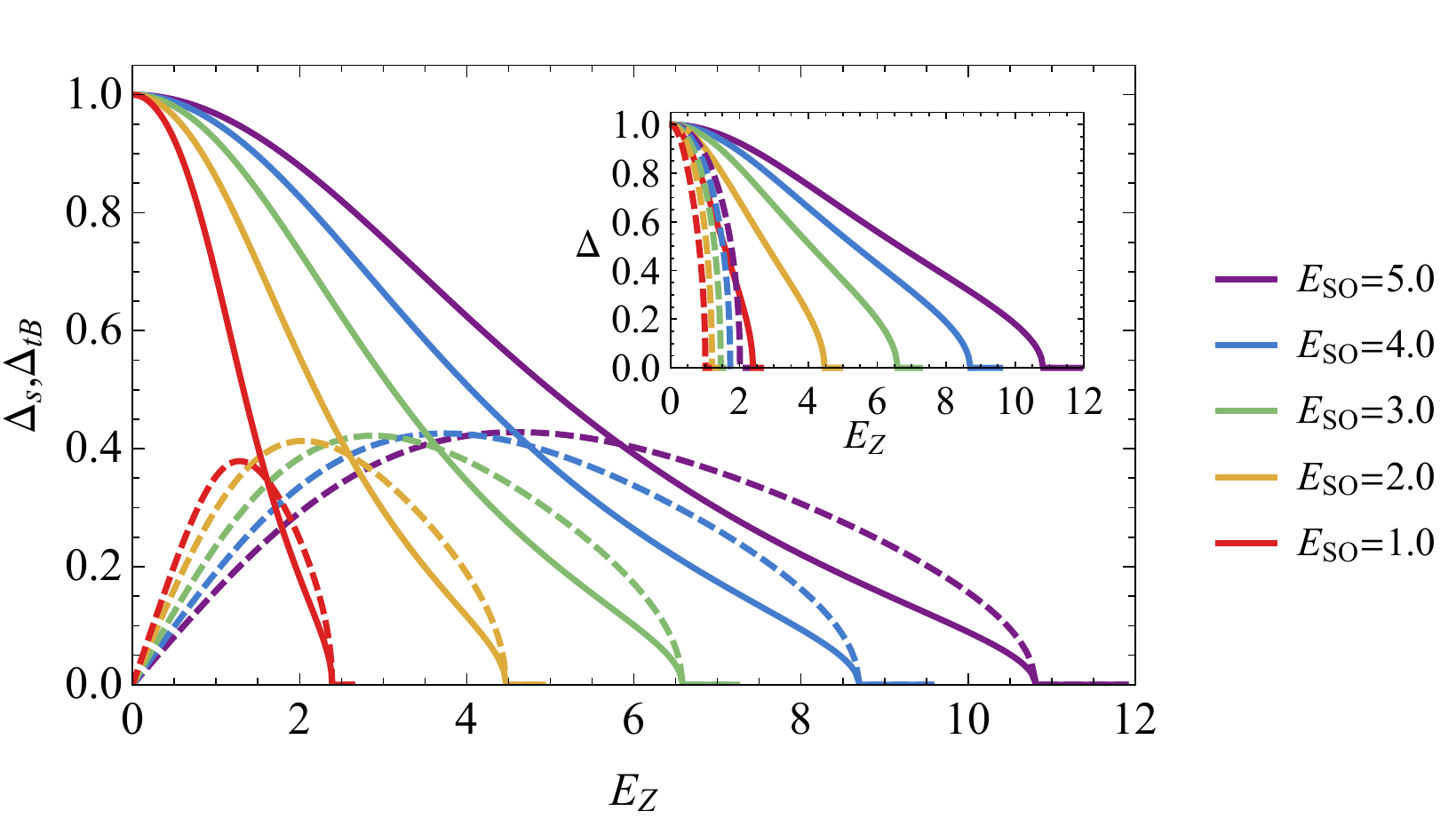}
\caption{\label{Fig4} Effect of the magnitude of the spin-orbit coupling: The singlet $\Delta_s$ (solid lines) and triplet $\Delta_{tB}$
(dashed lines) OPs as a function of the field for different values
of $E_{\mathrm{SO}}$ with parameters $T=0.75T_{cs},T_{ct}=0.7T_{cs},\tau_{iv}^{-1}=0.001T_{cs}$.
The inset shows the superconducting gap $\Delta$ for different values
of $E_{\mathrm{SO}}$ and with the same parameters as the main
graph, the solid lines are with the triplet component and the dashed
lines are for a singlet-only superconductor ($T_{ct}=0$). $E_Z,E_{\mathrm{SO}}$
are in units of $T_{cs}$. $\Delta,\Delta_s,\Delta_{tB}$
are normalized to the value of $\Delta=\Delta_s$ at $E_Z=0$.}
\end{figure}

\begin{figure}
\includegraphics[scale=0.51]{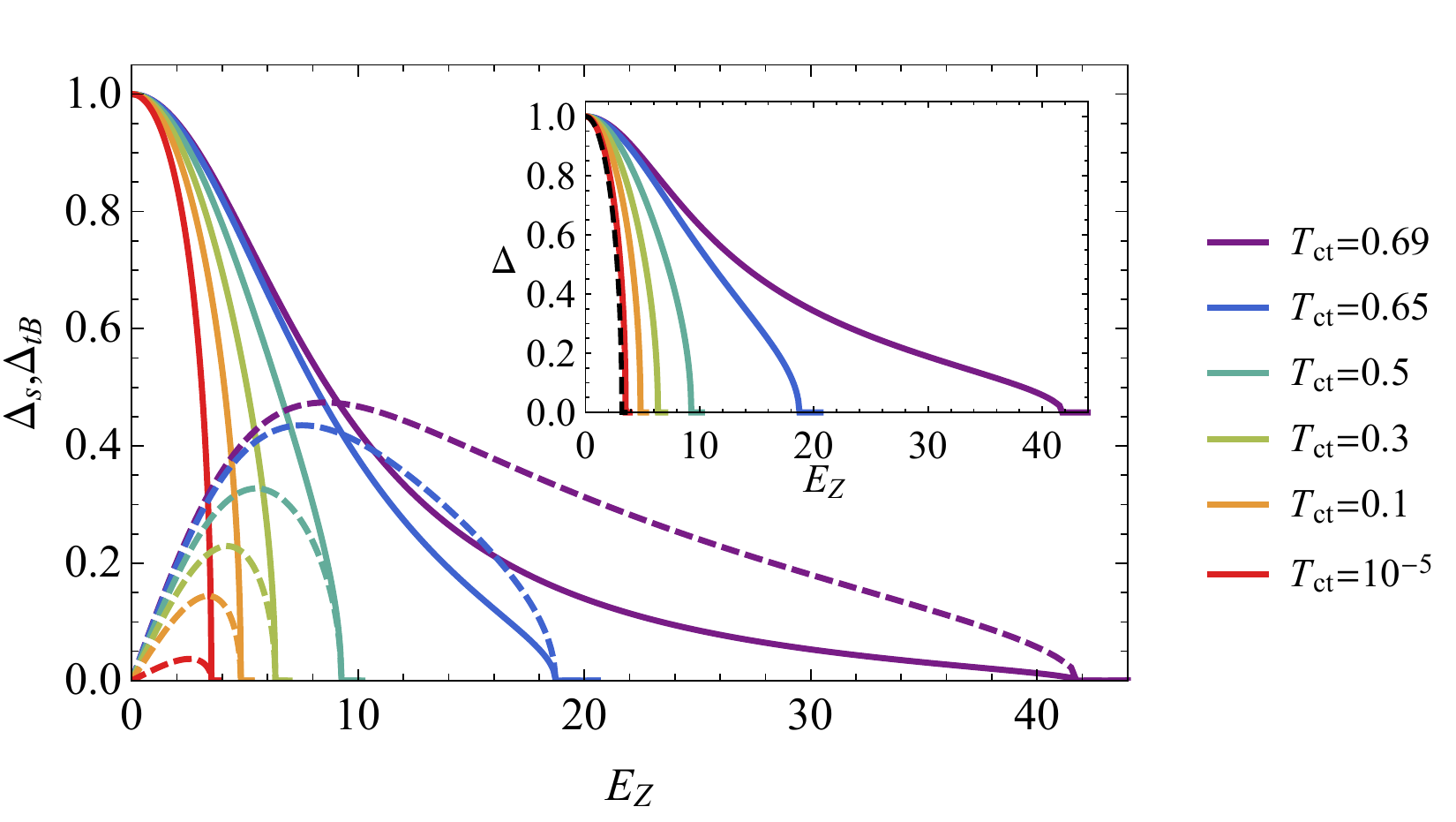}
\caption{\label{Fig5} Effect of the triplet pairing: The singlet $\Delta_s$ (solid lines) and triplet $\Delta_{tB}$
(dashed lines) OPs as a function of the field for different values
of $T_{ct}$ with parameters $T=0.7T_{cs},E_{\mathrm{SO}}=8.0T_{cs},\tau_{iv}^{-1}=0$.
The inset shows the superconducting gap $\Delta$ for different values
of $T_{ct}$ and with the same parameters as the main graph.
The black dashed line is the superconducting gap for a singlet-only superconductor ($T_{ct}=0$). $E_Z,T_{ct}$ are in units of $T_{cs}$.
$\Delta,\Delta_s,\Delta_{tB}$ are normalized
to the value of $\Delta=\Delta_s$ at $E_Z=0$.}
\end{figure}

Using the Landau expansion, we can obtain a qualitative understanding of the way the different
parameters affect  $\Delta_s$, $\Delta_{tB}$ and $\Delta$ as a function of the field. We start with the case of negligible inter-valley scattering. In Fig.~\ref{Fig4}, we see that for large enough $E_{\mathrm{SO}}$
the effect of increasing $E_{\mathrm{SO}}$ is only to stretch the lines for larger
critical fields $E_{Zc}$ but otherwise keeping the shape of lines as
they are. In Fig.~\ref{Fig5} ,we see that the effect of increasing $T_{ct}$
is to obtain larger critical fields $E_{Zc}$ by increasing the triplet component
in the superconducting phase, specifically we see that for larger $T_{ct}$we
get a steeper rise of the triplet component at low fields.

\begin{figure}
\includegraphics[scale=0.51]{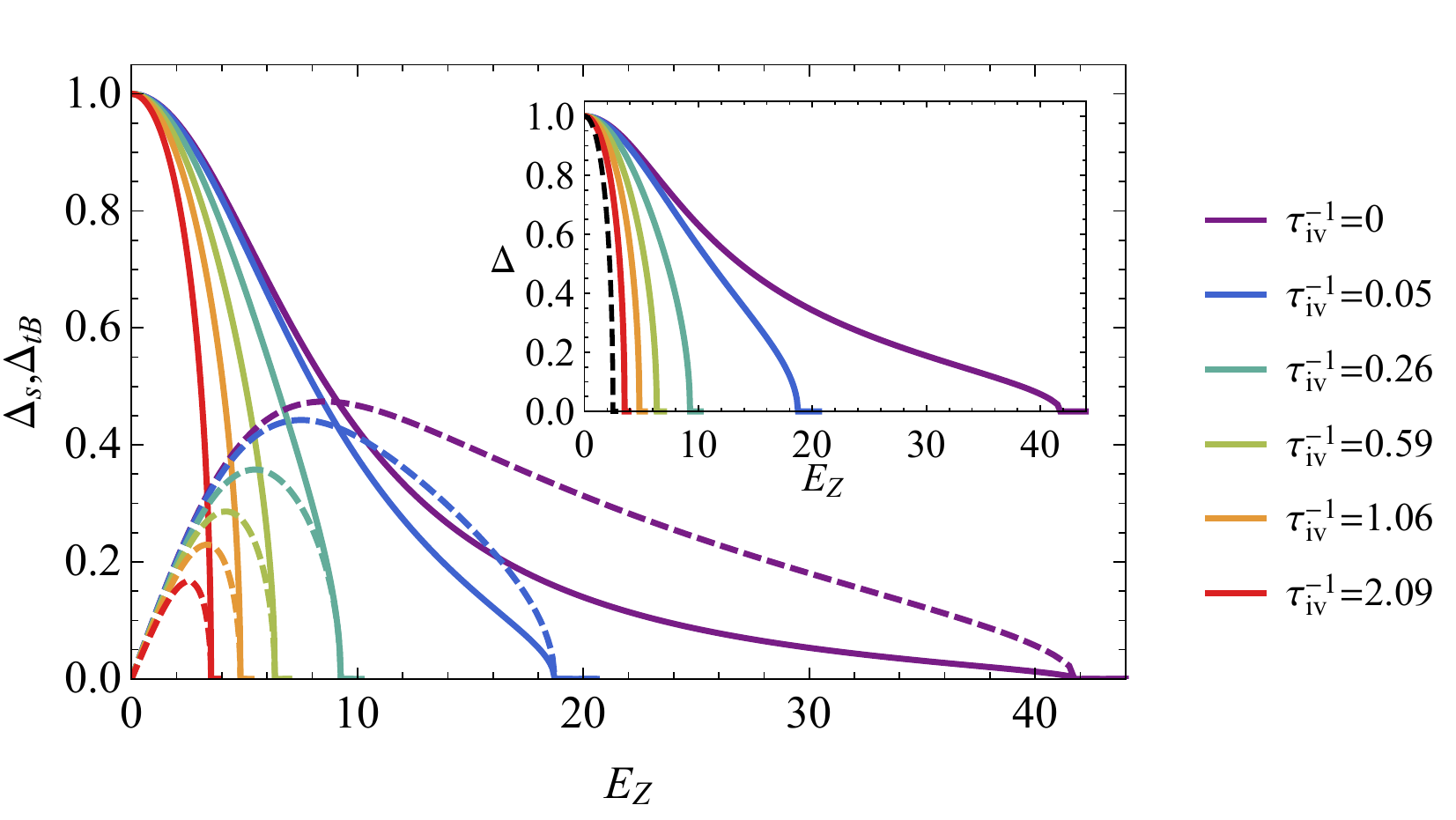}
\caption{\label{Fig6} Effect of disorder:  The singlet $\Delta_s$ (solid lines) and triplet $\Delta_{tB}$
(dashed lines) OPs as a function of the field for different values
of $\tau_{iv}^{-1}$ with parameters $T=0.7T_{cs},E_{\mathrm{SO}}=8.0T_{cs},T_{ct}=0.69T_{cs}$.
The inset shows the superconducting gap $\Delta$ for different values
of $\tau_{iv}^{-1}$ and with the same parameters as the main graph. The red {\tt use black as in Fig. S2 ?}
dashed line is the superconducting gap for a singlet-only superconductor ($T_{ct}=0$), and with $\tau_{iv}^{-1}=2.09T_{cs}$. $E_Z,\tau_{iv}^{-1}$ are in
units of $T_{cs}$. $\Delta,\Delta_s,\Delta_{tB}$
are normalized to the value of $\Delta=\Delta_s$ at $E_Z=0$.}
\end{figure}

We now turn to the effect of disorder. The impurity scattering potential has a broadening effect on the peak of the density of states but does not affect the form of the effective order parameter $\Delta$ appearing in the density of states significantly (though the superconducting energy gap may differ), hence we use \eqref{eq:1.22} as an estimation also in the presence of weak inter-valley scattering.
In the presence of the in-plane magnetic field, the scattering off the scalar impurities causes a spin flip with finite probability, and makes the scalar impurity to behave effectively as a magnetic impurity with a field-dependent concentration. While the problem is captured by the Abrikosov-Gor'kov theory of magnetic impurities~\cite{AG} in some parameter regimes, the general form of the self-consistency equation differs from the standard situation because the spin splitting $E_{\mathrm{SO}}$ intervenes as an additional energy scale.

The parameters corresponding to the lines with the highest critical fields in Figs.~\ref{Fig5} and \ref{Fig6} are identical. The $\tau_{iv}^{-1}$ parameters in Fig.~\ref{Fig6} were chosen so that identical colors in Figs.~\ref{Fig5} and \ref{Fig6} will have approximately the same critical field.   
In Fig.~\ref{Fig6}
we see that by increasing $\tau_{iv}^{-1}$ we negate the affect of
having $T_{ct}>0$. Comparison of the OPs suppression obtained in  Fig.~\ref{Fig5} by decreasing $T_{ct}$ to the suppression obtained in Fig.~\ref{Fig6} by increasing $\tau_{iv}^{-1}$ shows that in the latter process we can retain a relatively steep increase of the triplet OP even for small critical fields, while in Fig.~\ref{Fig5} the decrease in the critical field is accompanied by a faster decrease in slope of the triplet. The reason for this is that, even though increasing $\tau_{iv}^{-1}$ in Fig.~\ref{Fig6} suppresses superconductivity, we still keep a high $T_{ct}$, which strengthens the triplet component, while in  Fig.~\ref{Fig5} the suppression of superconductivity is achieved by direct suppression of the triplet component. The gap contains contributions of both the triplet and singlet order parameters. As the triplet order parameter is affected more strongly by a suppression of $T_{ct}$ than by increase in $\tau_{iv}^{-1}$ the same is true for the gap.
Compared to the triplet order parameter taken separately, the distinction between $\Delta(H)$ in the two cases is less pronounced as long as the singlet order parameter makes a dominant contribution to the gap.

\subsection{Two-pocket superconductivity in NbSe$_2$ - Suhl-Matthias-Walker coupling}\label{sec:two-pockets-MSW}

In few-layer $\mathrm{NbSe_2}$ the $\mathrm{Se}$ derived Fermi pockets, observed in the bulk~\cite{noat2015quasiparticle}, disappear~\cite{Wickramaratne2020}, and only the $\mathrm{Nb}$ derived pockets close to the $K/K'$ and $\Gamma$ points are left. Here we discuss the multi-pocket effects assuming a Suhl-Matthias-Walker\cite{Suhl1959} type coupling between pockets.
This coupling describes the the inter-pocket  Cooper pair tunneling process.

\subsubsection{OPs in the two-pocket model at zero field, \texorpdfstring{$E_Z =0$}{TEXT}}
\label{sec:OPs_ZeroB}

The purpose of this section is to elucidate the limitations on the parameters of the two-pocket model imposed by the zero field data (cf. inset of Figure~\ref{fig:GvsT}), where the BCS ratio of the superconducting gap to the critical temperature is found. This implies that, in the case where there are two superconducting gaps, the larger one is measured in the experiment. In addition, in the experiment, quasiparticles tunnel into NbSe$_2$ mainly at the K/K' points, as this is where the energy gap is the smallest in the tunnel barrier material (few-layer WSe$_2$ or MoS$_2$)~\cite{zhu,wang2012}. In the previous work of some of us on bulk NbSe$_2$ with similar barriers, we found a negligible contribution from tunneling into the $\Gamma$ point~\cite{Dvir2017}. Therefore, we consider only tunneling into the K/K' points. 

\begin{figure*}
\centering
\includegraphics[width=0.7\textwidth]{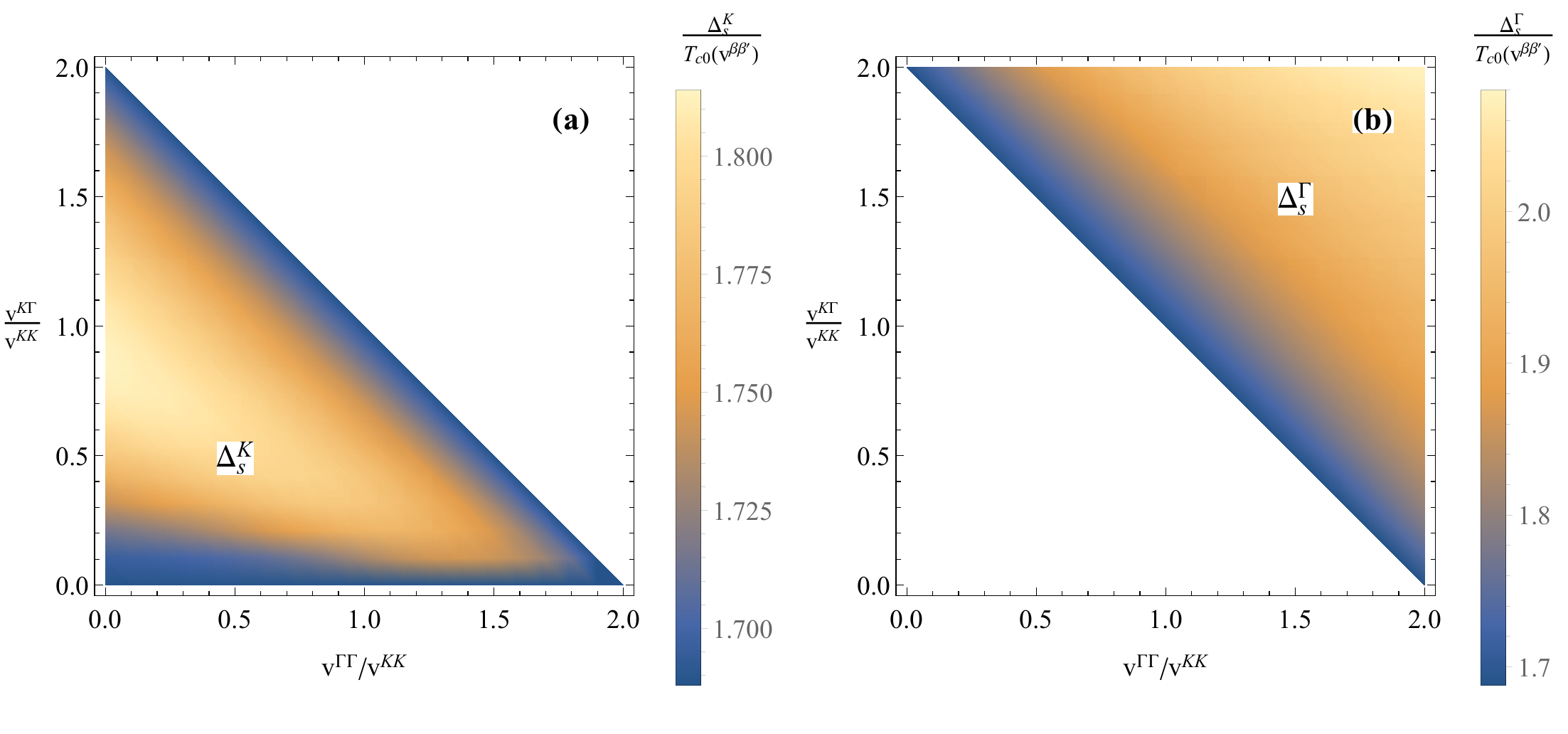}
\caption{\label{fig:res1}The OPs at $E_{Z}=0$ for different values of $v^{\Gamma\Gamma}/v^{KK}$
and $v^{K\Gamma}/v^{KK}$, as calculated by numerically
solving the two self-consistency equations. 
The plots show the larger OP, which is $\Delta_{s}^{K}$ in the region of panel $\mathbf{\left(a\right)}$ and $\Delta_{s}^{\Gamma}$ in the region of panel $\mathbf{\left(b\right)}$. We take $N_{0}v^{KK}=-0.5$, and $T/T_{c0}=0.5$ where $T_{c0}$ is given by Eq.~\eqref{eq:Tc0_v} and
the OPs is in units of $T_{c0}$.
}
\end{figure*}

Specifically, we find that the BCS ratio of the superconducting gap to the critical temperature is achieved either when the $\Gamma$ and $K/K'$ pockets are decoupled or when the interaction amplitudes satisfy the following condition, 
\begin{align}\label{EQ:CONDITION}
    v^{\Gamma\Gamma}+ 2v^{K\Gamma}=2v^{KK} + v^{K\Gamma}\, .
\end{align}
The decoupled pockets are analyzed above, and here we consider the coupled pockets under the condition, \eqref{EQ:CONDITION}.
This relation follows as the number of $K$ pockets is twice as large than the number of $\Gamma$ pockets.
The relation \eqref{EQ:CONDITION} is confirmed numerically in the Fig.~\ref{fig:res1}.
where the ratio of the gap $\Delta_s^K$ to the zero field critical temperature,
\begin{equation}
\label{eq:Tc0_v}
T_{c0}\left(v^{\beta\beta'}\right)=\frac{2\omega_{D}e^{\gamma_{E}}}{\pi}\exp\left(-\frac{v^{KK}+v^{\Gamma\Gamma}/2+\sqrt{2\left(v^{K\Gamma}\right)^{2}+\left(v^{KK}-v^{\Gamma\Gamma}/2\right)^{2}}}{4N_{0}\left[\left(v^{K\Gamma}\right)^{2}-v^{\Gamma\Gamma}v^{KK}\right]}\right).
\end{equation}
is shown as a function of $v^{\Gamma\Gamma}/v^{KK}$ and $v^{K\Gamma}/vv^{KK}$. As the experimentally observed ratio is close to the BCS ratio, we restrict ourselves in the following to this line, which includes, in particular, the case where all interactions are equal.

\subsubsection{The case of equal inter- and intra-pocket interactions} \label{sec:equal_amp}

We now consider the case $v^{\Gamma\Gamma}=v^{KK}=v^{K\Gamma}$. 
In this section we show that this case is equivalent to a single pocket. 
It can be shown from the self consistency equations that in this case the OPs are equal $\Delta_{s}^{K}=\Delta_{s}^{\Gamma}=\Delta_{s}$,
and the two self-consistency equations reduce to one:
\begin{equation}
\label{eq:self_lambda_equal_nu}
\pi T\underset{\omega_{n}}{\sum}\left[\frac{\Delta_{s}}{\left|\omega_{n}\right|}-\left(\frac{2}{3}\left\langle f_{0}^{K}\left(\hat{\mathbf{k}},\omega_{n},\Delta_{s}\right)\right\rangle _{\varphi_{\hat{\mathbf{k}}}}+\frac{1}{3}\left\langle f_{0}^{\Gamma}\left(\hat{\mathbf{k}},\omega_{n},\Delta_{s}\right)\right\rangle _{\varphi_{\hat{\mathbf{k}}}}\right)\right]+\Delta_{s}\ln\left(\frac{T}{T_{c0}}\right) =0\, ,
\end{equation}
where $f_{0}^{\Gamma(K)}\left(\hat{\mathbf{k}},\omega_{n},\Delta_{s}\right)$ are quasi-classical Green functions, \cite{haim2020signatures}. 
That is, we effectively have a single pocket with field
dependence which averages over the pair breaking effect at the $K/K'$ and at $\Gamma$ pockets.

The physical picture contained in Eq.~\eqref{eq:self_lambda_equal_nu} is that of a single band, albeit anisotropic.
For this reason, it has been possible to parametrize Eq.~\eqref{eq:self_lambda_equal_nu} by the critical temperature, $T_{c0}$ instead of the pairing amplitude $\lambda$.
In general, this is not possible even close to $T_{c0}$ \cite{Kogan2011}.
Furthermore, the pair breaking equation for the critical field reads,  
\begin{equation}
\label{eq:pair_breaking_lambda_equal_nu}
\ln\left(\frac{T}{T_{c0}}\right)+\left(\frac{2}{3}\mathcal{S}_{K}+\frac{1}{3}\mathcal{S}_{\Gamma}\right)=0\, , 
\end{equation}
where
\begin{equation}
\label{eq:SKG}
\mathcal{S}_{K}=\pi T\underset{\omega_{n}}{\sum}\frac{1}{\left|\omega_{n}\right|}\left(\frac{E_{Z}^{2}}{E_{Z}^{2}+\left(E_{\mathrm{SO}}^{K}\right)^{2}+\omega_{n}^{2}}\right)\, ,\quad 
\mathcal{S}_{\Gamma}=\pi T\underset{\omega_{n}}{\sum}\frac{1}{\left|\omega_{n}\right|}\left(\frac{E_{Z}^{2}}{\sqrt{\left(E_{Z}^{2}+\omega_{n}^{2}\right)\left(E_{Z}^{2}+\left(E_{\mathrm{SO}}^{\Gamma}\right)^{2}+\omega_{n}^{2}\right)}}\right).
\end{equation}
The last equation can be obtained by linearizing Eq.~\eqref{eq:self_lambda_equal_nu}.
Again, it evidently represents the average over the combined Fermi surface made out of $K/K'$ and $\Gamma$ pockets.

\subsubsection{The effect of impurities on the OP}
\label{sec:Impurities}
In the case where all the interactions amplitudes are equal, we saw
that we effectively have a single band comprised of pairing in the
$K/K'$ and $\Gamma$ pockets. Neglecting inter-valley scattering, the presence of impurities can effect
the OP through the $\Gamma$ pairing. This is because the spin-orbit nodes in the $\Gamma$ pocket lead to the 'flattened' shape of the OP as a function of in-plane magnetic field at intermediate fields. (A model with two nodeless pockets cannot reproduce this aspect of the data.)

\begin{figure*}
\centering
\includegraphics[width=0.9\textwidth]{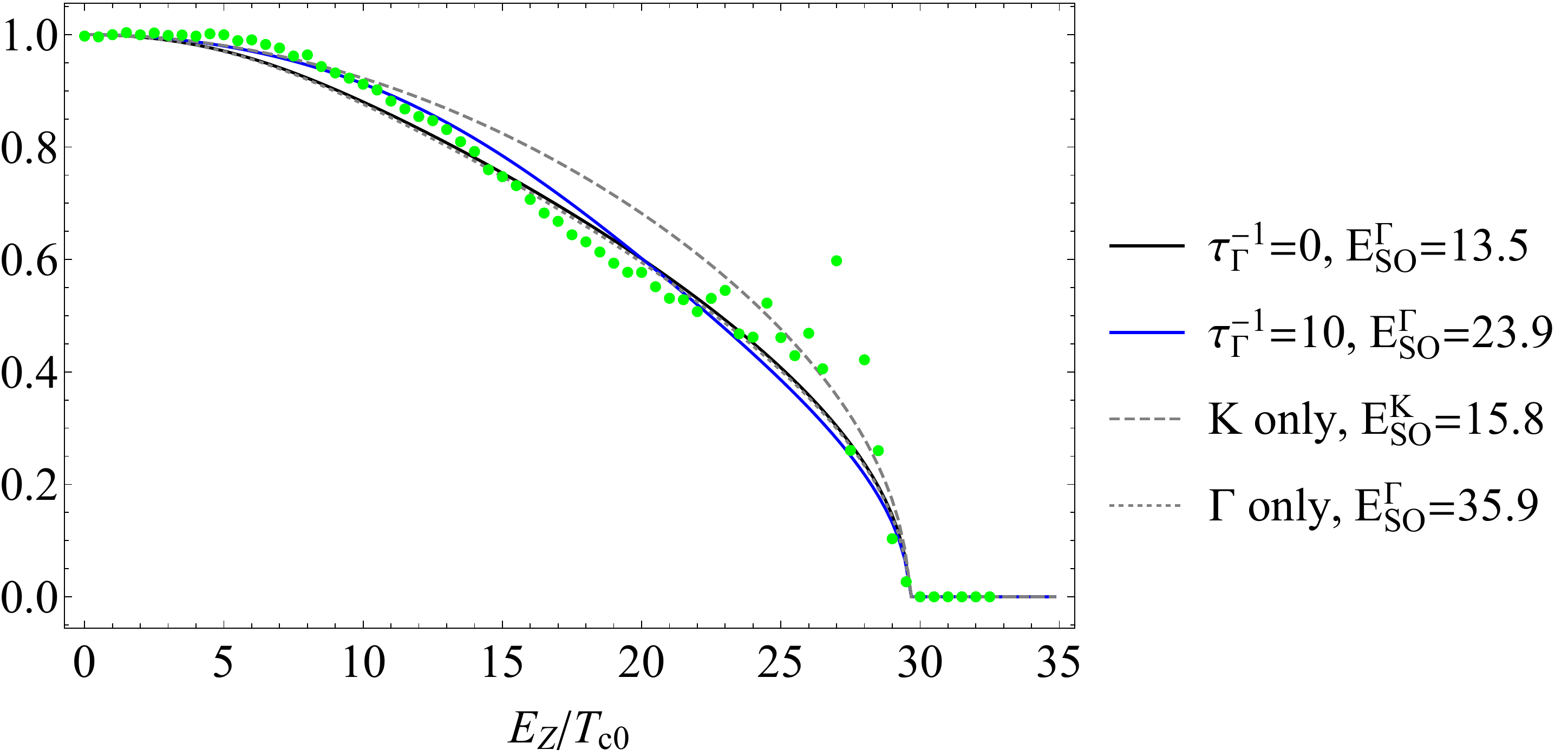}
\caption{\label{fig:res4} 
The OP as a function of $E_Z$. The black and blue lines have $K$ and $\Gamma$ equal pairing interactions  and are calculated by numerically solving Eq.~ \eqref{eq:self_lambda_equal_nu} for the clean case (black line) and with a high impurity scattering rate of $\tau_{\Gamma}^{-1} = 10$ (blue line). The dashed and dotted lines show the OP in the clean case with only the $K$ and $\Gamma$ pocket pairings, respectively. 
These are found by replacing in Eq.~\eqref{eq:self_lambda_equal_nu} $f_{0}^{\Gamma}$ by $f_{0}^{K}$ for the $K$ pocket and replacing $f_{0}^{K}$ with the $f_{0}^{\Gamma}$ for the $\Gamma$ pocket. 
The green dots show the experimental data and the SOC for each line is taken to fit the experimental critical field. 
The parameters  $\tau^{-1}_{\Gamma}$ and $E_{\mathrm{SO}}^{\Gamma},E_{\mathrm{SO}}^{K}$ are in units of $T_{c0}$ and for the black and blue lines we take $E_{\mathrm{SO}}^{K}/E_{\mathrm{SO}}^{\Gamma}=2.14$. 
}
\end{figure*}

In order to include this effect, we write the Gorkov equation as
\begin{equation}
\label{eq:Gorkov_imp}
\left[i\omega_{n}\hat{\sigma}_{0}-\hat{H}_{\mathrm{BdG}}^{\Gamma}\left(\mathbf{k}\right)-\hat{\Sigma}^{\left(\Gamma\right)}\left(\omega_{n}\right)\right]\hat{G}^{\Gamma}\left(\mathbf{k},\omega_{n}\right)=\hat{\sigma}_{0}
\end{equation}
where $\hat{\Sigma}^{\left(\Gamma\right)}\left(\omega_{n}\right)$
is the self-energy given withing the self-consistent Born approximation
by \cite{Wickramaratne2021}
\begin{equation}
\label{eq:self_energy}
\hat{\Sigma}^{\left(\Gamma\right)}\left(\omega_{n}\right)=\tau^{-1}_{\Gamma}\int\frac{d\varphi_{\mathbf{k}}}{2\pi}\int\frac{d\xi_{\mathbf{k}}^{\Gamma}}{\pi}\hat{\sigma}_{z}\hat{G}^\Gamma\left(\mathbf{k};\omega_{n}\right)\hat{\sigma}_{z}\, ,
\end{equation}
where $\hat{\sigma}_{z}=\mathrm{diag}\left(\sigma_{0},-\sigma_{0}\right)$
and $\tau^{-1}_{\Gamma}$ is intra $\Gamma$ pocket impurity
scattering rate. We calculate the OP by finding the root of the self
consistency \eqref{eq:self_lambda_equal_nu}, where $f_{0}^{\Gamma}\left(\hat{\mathbf{k}},\omega_{n}\right)$
now includes the effect of impurities as in Eq.~\eqref{eq:Gorkov_imp}.
In order to find the root we must evaluate $\left\langle f_{0}^{\Gamma}\left(\hat{\mathbf{k}},\omega_{n}\right)\right\rangle_{\varphi_{\hat{\mathbf{k}}}}$ for
any given $\Delta_{s}$. This is done by calculating $\hat{\Sigma}^{\left(\Gamma\right)}\left(\omega_{n},\Delta_{s}\right)$
for the given $\Delta_{s}$ by way of iterations using Eqs.~(\ref{eq:Gorkov_imp},\ref{eq:self_energy}),
then inserting $\hat{\Sigma}^{\left(\Gamma\right)}\left(\omega_{n},\Delta_{s}\right)$
into Eq.~\eqref{eq:Gorkov_imp} and finding $\left\langle f_{0}^{\Gamma}\left(\hat{\mathbf{k}},\omega_{n}\right)\right\rangle_{\varphi_{\hat{\mathbf{k}}}} $.
The results for different values
of $\tau^{-1}_{\Gamma}$ are presented in Fig.~\ref{fig:res4}. In our numerical calculations we take $E_{\mathrm{SO}}^{K}/E_{\mathrm{SO}}^{\Gamma}=2.14$ based on  Ref.~\cite{Wickramaratne2020} assuming that this ratio is roughly similar for mono- and bilayer systems.

Without disorder a reasonable fit to the experimental data is achieved (Figure~\ref{fig:res4}). The level of experimental disorder is estimated to be at least $\approx$ 10 meV (cf. Section IC), which is to say tens of $T_{cs}$. The inclusion of such a strong disorder at the level of the self-consistent Born approximation makes the procedure numerically unstable. 
Nevertheless, even at the levels of disorder we are able to incorporate, it can already be seen that the fit becomes progressively worse with increasing disorder (Figure~\ref{fig:res4}). At even higher values of disorder scattering rate, we expect it to continue to get worse compared to the one obtained with the triplet pairing. 
In addition, with increasing disorder, an increase in the value of spin-orbit coupling necessary to obtain the same critical field. Finally, at very strong levels of disorder, i.e. $1/\tau_{\Gamma} \gg \Delta_{SO}^\Gamma$, the $\Gamma$-pocket is described by AG-theory which yields a field-dependence of the OP inconsistent with the data.

\FloatBarrier

\subsection{Two-pocket superconductivity in NbSe$_2$ - McMillan coupling}
\label{sec-mm}

Here we address the possibility that multipocket physics, with a McMillan coupling\cite{mcmillan} plays a significant role in our findings.
As before we, consider the $K(K')$ and $\Gamma$ pockets, originating from the $\mathrm{Nb}$ derived band\cite{Wickramaratne2020}.
Unlike in section \ref{sec:two-pockets-MSW} where two-particle intra-pocket tunneling was discussed, the McMillan coupling considered here amounts to single-particle tunneling.

The theoretical model is presented in section \ref{sec:mcmillan_model},  a comparison with experimental data is given in section \ref{sec:mcmillan_exp}, followed by a short discussion in section \ref{sec:mcmillan_comments}.

\subsubsection{Multipocket model}\label{sec:mcmillan_model}

We account for scattering within and between pockets by the following rates:
\begin{itemize}
	\item $\tau^{-1}$: scattering within the $K,K'$ pockets (intravalley scattering). Note that this type of disorder has no effect on superconductivity due to the Anderson theorem,
	\item $\tau_{iv}^{-1}$: Scattering between $K$ and $K'$ pockets (intervalley scattering),
	\item $\tau_\Gamma^{-1}$: scattering within the $\Gamma$ pocket, 
	\item $\Gamma_{K\Gamma}$, $\Gamma_{\Gamma K}$: scattering from the $K$ to the $\Gamma$ pocket and vice-versa. 
\end{itemize}
The following relation holds for interpocket scattering rates
\begin{equation}
\frac{\Gamma_{K\Gamma}}{\Gamma_{\Gamma K}}=\frac{N_\Gamma}{N_K},
\label{eq2}
\end{equation}
where $N_K$ is the sum of the normal state DoS of the K and K' pockets, and $N_\Gamma$ is the normal state DoS of the $\Gamma$ pocket. Taking into account that DoS per pocket is approximately the same for $K$, $K'$ and $\Gamma$, we have $N_K\approx 2 N_\Gamma$ \cite{sticlet2019topological}. 

We will assume $s$-wave superconductivity, that originates from the $K$ pocket, and is induced by the proximity effect in the $\Gamma$ pocket through interpocket scattering. Furthermore, we will assume diffusive limit in the $\Gamma$ pocket, $\tau_\Gamma^{-1}\gg E_{SO}^\Gamma, E_Z, \Delta_0, \Gamma_{ij}$. If the SOC is sufficiently strong in both pockets, that is, if $E_{SO}^K\gg \Delta_0, \Gamma_{ij}, E_Z, \tau_{iv}^{-1}$, and $(E_{SO}^\Gamma)^2\tau_\Gamma \gg \Delta_0,E_Z,\Gamma_{ij}$, the density of states is given by the Kaiser-Zuckermann formula \cite{kaiser1970mcmillan}, which is a multiband extension of the Abrikosov-Gor'kov theory.  Namely, depairing is both $K$ and $\Gamma$ pockets is captured by the rates quadratic in magnetic field
\begin{equation}
\Gamma_{AG}^K(E_Z)=\frac{E_Z^2}{(E_{SO}^K)^2 \tau_{iv}}, \qquad
\Gamma_{AG}^\Gamma(E_Z)=\frac{E_Z^2}{(E_{SO}^\Gamma)^2 \tau_\Gamma},
\end{equation}
and we can introduce the functions $u_K$ and $u_\Gamma$ that satisfy
\begin{equation}
\frac{\omega_n}{\Delta}=u_K-\frac{\Gamma_{AG}^K(E_Z)}{\Delta}\frac{u_K}{\sqrt{1+u_K^2}}+\frac{\Gamma_{K\Gamma}}{\Delta}\frac{u_K-u_\Gamma}{\sqrt{1+u_\Gamma^2}} = -\frac{\Gamma_{AG}^\Gamma(E_Z)}{\Delta}\frac{u_\Gamma}{\sqrt{1+u_\Gamma^2}}+\frac{\Gamma_{\Gamma K}}{\Delta}\frac{u_\Gamma-u_K}{\sqrt{1+u_K^2}}.
\label{eq4}
\end{equation}
Here, $\Delta$ is the intrinsic gap from the $K$ pocket. Then, the DoS in the superconducting state is given as
\begin{equation}
\nu(E)=\sum_{i=K,\Gamma} N_i \Re\left[\frac{u_i}{\sqrt{1+u_i^2}}\right]_{i\omega_n\to E}.
\end{equation}

The intrinsic gap $\Delta$ satisfies the self-consistency condition
\begin{equation}
\Delta=2\pi T N_K |v_s| \sum_{\omega_n>0}\frac{1}{\sqrt{1+u_K^2}}.
\label{eq6}
\end{equation}
An alternative formulation is given by Eq. \eqref{eqscalt}, parameterized in terms of the critical temperature $T_c$ instead of the coupling constant $v_s$
\begin{equation}
\ln \bigg(\frac{T}{T_c}\bigg)+\frac{\Gamma_{K\Gamma}}{\Gamma_{K\Gamma}+\Gamma_{\Gamma K}}\bigg[\psi \bigg(\frac{1}{2}\bigg)-\psi \bigg(\frac{1}{2}+\frac{\Gamma_{K\Gamma}+\Gamma_{\Gamma K}}{2 \pi  T_c}\bigg)\bigg]=2 \pi T \sum_{\omega_n>0} \bigg[\frac{1}{\Delta}\frac{1}{\sqrt{1+u_K^2}}  - \frac{1}{\omega_n} \bigg].
\label{eqscalt}
\end{equation}
Here, $\psi(x)$ if the digamma function

Close to the phase transition to the normal state, that is, in the limit of  vanishing $\Delta$, it is possible to analytically evaluate Eq. \eqref{eq6}. The self-consistency condition then becomes
\begin{equation}
\ln \bigg(\frac{T}{T_c}\bigg)+\frac{\Gamma_{K\Gamma}}{\Gamma_{K\Gamma}+\Gamma_{\Gamma K}}\bigg[\psi \bigg(\frac{1}{2}\bigg)-\psi \bigg(\frac{1}{2}+\frac{\Gamma_{K\Gamma}+\Gamma_{\Gamma K}}{2 \pi  T_c}\bigg)\bigg]= \sum_{i=\pm} \frac{\alpha_i}{\chi^2}\bigg[
\psi \bigg(\frac{1}{2}+\frac{\beta_i}{2\pi  T}\bigg)-\psi\bigg(\frac{1}{2}\bigg) \bigg)
\bigg].
\label{eq7}
\end{equation}  
Here $T_c$ is the critical temperature. Furthermore, we have introduced  
\begin{align}
\chi= &\,\bigg(4 \Gamma_{K\Gamma}\Gamma_{\Gamma K}+[\Gamma_{K\Gamma}-\Gamma_{\Gamma K}+\Gamma_{AG}^K(E_{Z}^{c})-\Gamma_{AG}^\Gamma(E_Z^{c})]^2\bigg)^{1/2}, \nonumber \\
\alpha_{\pm}= &\, \frac{1}{2}\bigg(-\chi^2\pm\chi [\Gamma_{K\Gamma}-\Gamma_{\Gamma K}+\Gamma_{AG}^K(E_{Z}^c)-\Gamma_{AG}^\Gamma(E_{z}^c)]\bigg), \nonumber  \\
\beta_{\pm}= & \,\frac{1}{2}\bigg(\Gamma_{K\Gamma}+\Gamma_{\Gamma K}+\Gamma_{AG}^K(E_Z^c)+\Gamma_{AG}^\Gamma(E_Z^c)\mp \chi \bigg),
\end{align}
where $E_Z^c$  is the upper critical field. Equation \eqref{eq7} can be used to calculate the $E_Z^c$, provided that the scattering parameters are known. 

\subsubsection{Comparison with the experiment}\label{sec:mcmillan_exp}

The main goal is to obtain the $\Delta(H)$ dependence, based on equation \eqref{eqscalt}. As it stands there are five parameters: the critical temperature $T_c$, the scattering rates $\Gamma_{K\Gamma}$ and $\Gamma_{\Gamma K}$, and the Abrikosov-Gorkov depairing parameters $\Gamma_{AG}^K$ and $\Gamma_{AG}^\Gamma$.
In order to reduce the size of the parameter space, and to ease the exploration, some physically motivated constraints were placed on the parameter values.\\

\subsubsubsection{Zero magnetic field}

At zero field the Abrikosov-Gorkov parameters do not play a role and the order parameter is a function of $\Delta = \Delta(T,T_c,\Gamma_{K\Gamma},\Gamma_{\Gamma K})$.
The temperature was fixed to the experimental one ($T=1.3\mathrm{K}$) and the ratio of the pocket scattering rates is set according to Eq. \eqref{eq2}: $\Gamma_{\Gamma K} = 2\Gamma_{K\Gamma}$. 

\begin{figure}[h!]
	\centering
	\includegraphics[width=0.5\textwidth]{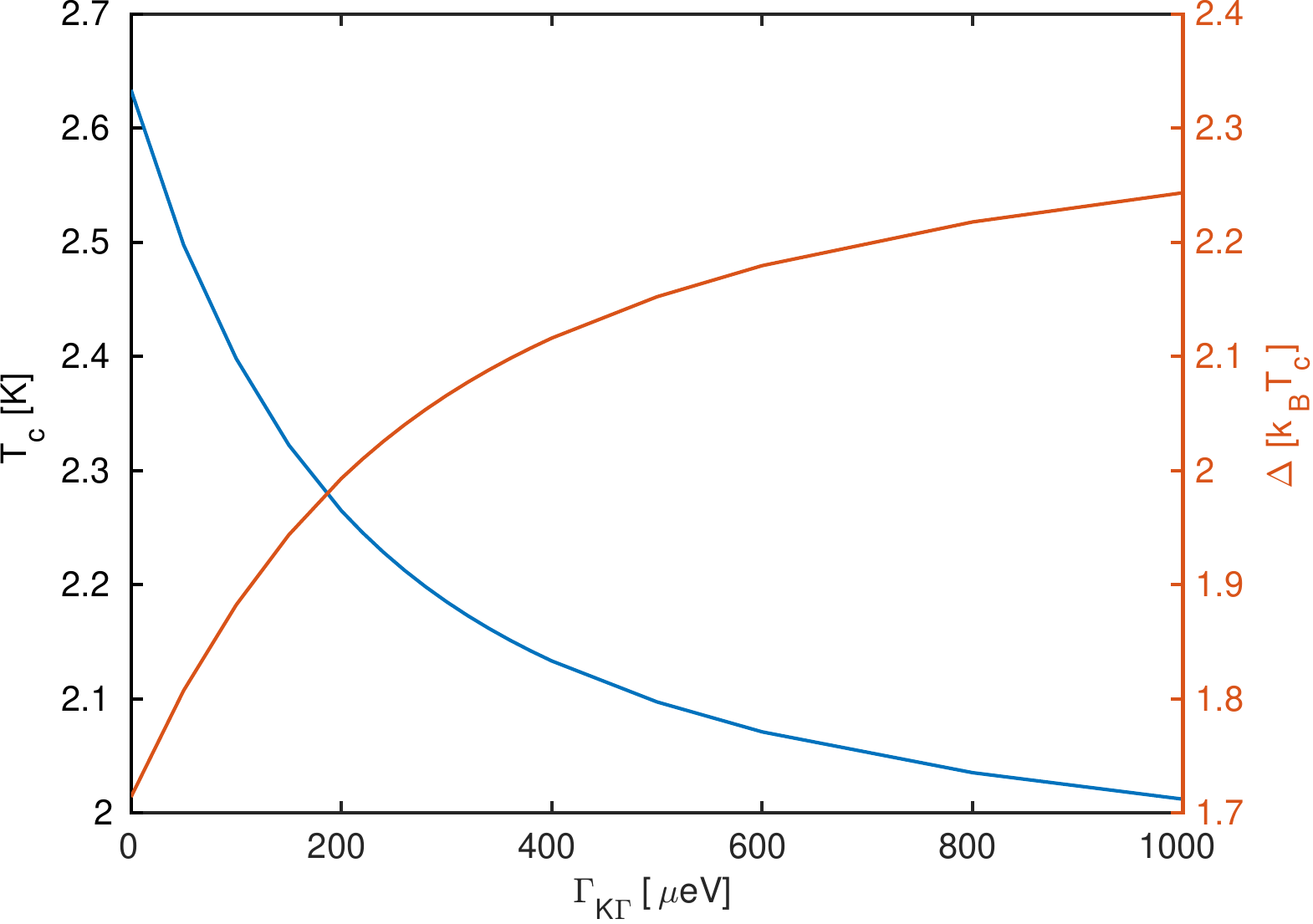}
	\caption{The dependence of $T_c$ on $\Gamma_{K\Gamma}$, constrained to $\Delta\approx 390\mathrm{\mu eV}$ at $T=1.25\mathrm{K}$.}
	\label{fig:TcvsGamma}
\end{figure}

\begin{figure}[!htb]
	\centering
	\begin{minipage}{.48\textwidth}
		\centering
		\includegraphics[width=\textwidth]{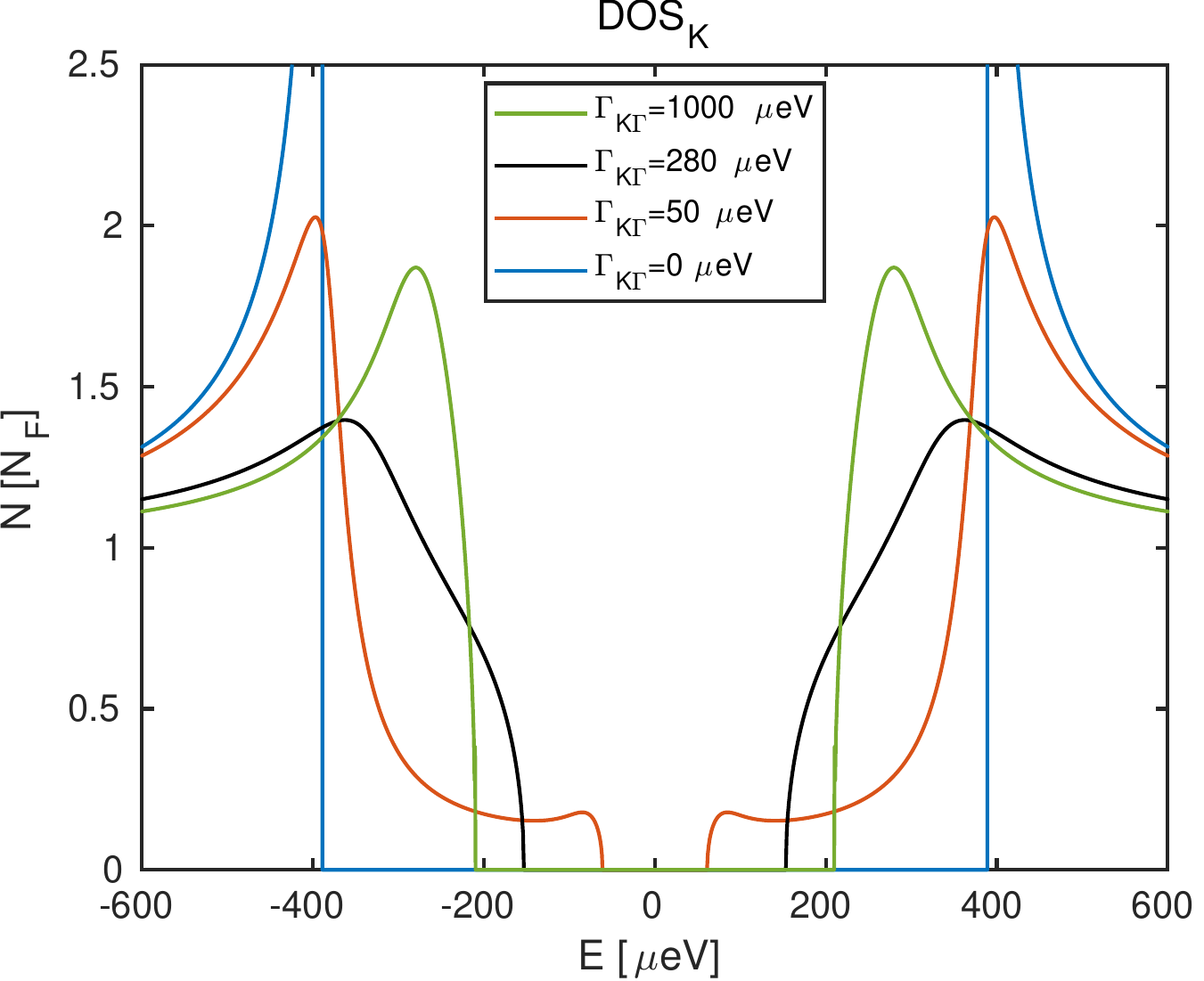}
		\caption{The DOS in the $K$ pocket vs $\Gamma_{K\Gamma}$.}
		\label{fig:dosK}
	\end{minipage}%
	\begin{minipage}{0.48\textwidth}
		\centering
		\includegraphics[width=\textwidth]{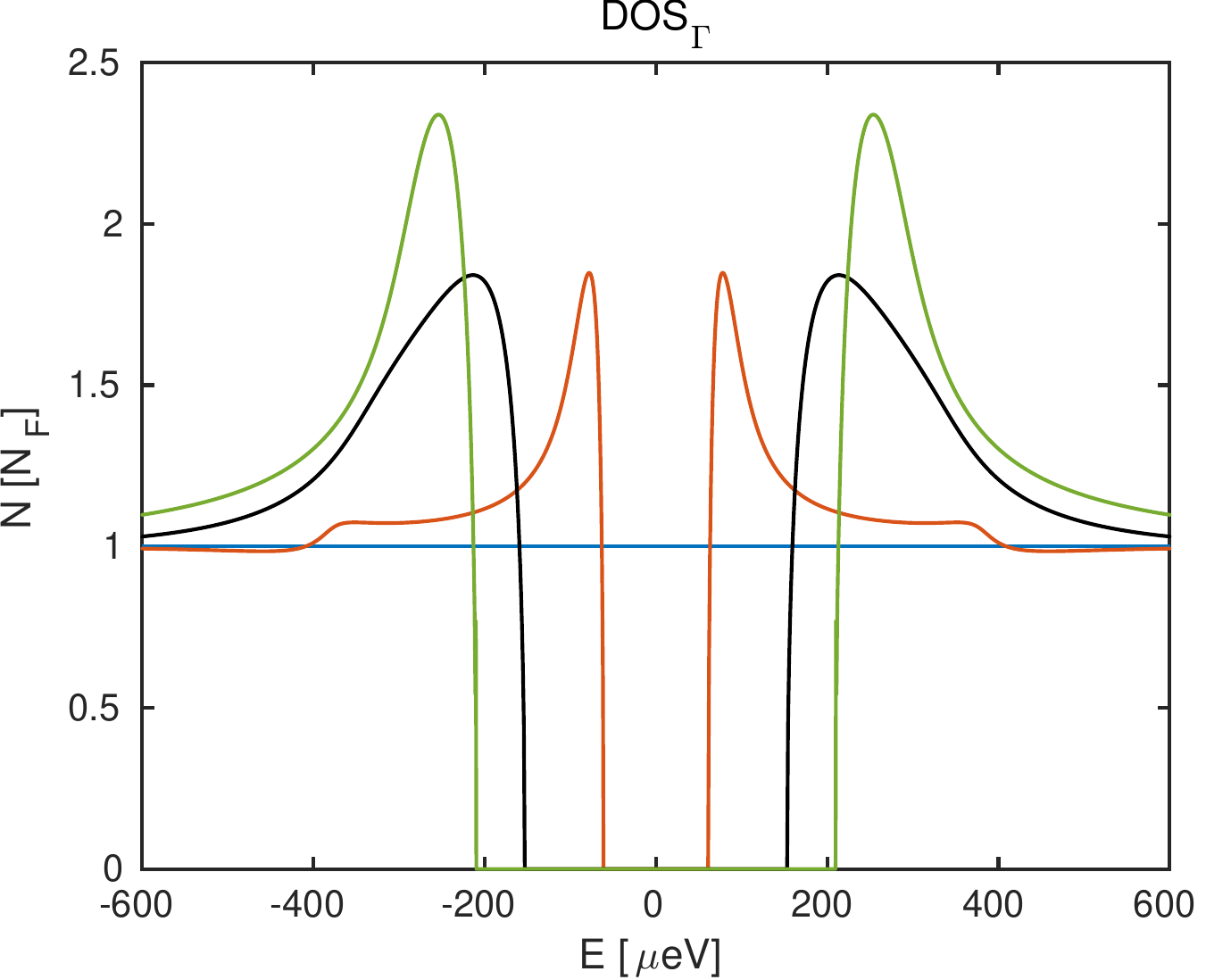}
		\caption{The DOS in the $\Gamma$ pocket vs $\Gamma_{K\Gamma}$.}
		\label{fig:dosG}
	\end{minipage}
\end{figure}

With this, at $H=0$ it remains to determine $T_c$, this was done by constraining the value of $\Delta$ to $\approx 390\mathrm{\mu eV}$, as obtained in the experiment  (cf. inset of Figure~\ref{fig:GvsT}). Then we can solve for $T_c$ as a function of $\Gamma_{K\Gamma}$. The result is shown on figure \ref{fig:TcvsGamma}: at zero coupling the ratio $\frac{\Delta}{k_BT_c}$ starts from the BCS value of $\approx1.76$ and grows as the coupling is increased. Therefore for the same value of $\Delta$ the critical temperature of a multipocket superconductor is lower than that of a BCS one.
The densities of states of both pockets, at $H=0$, for select values of $\Gamma_{K\Gamma}$ are shown in figures \ref{fig:dosK} and \ref{fig:dosG}.
\newline

\subsubsubsection{Finite magnetic field}

\begin{figure}[h!]
	\centering
	\includegraphics[width=0.5\textwidth]{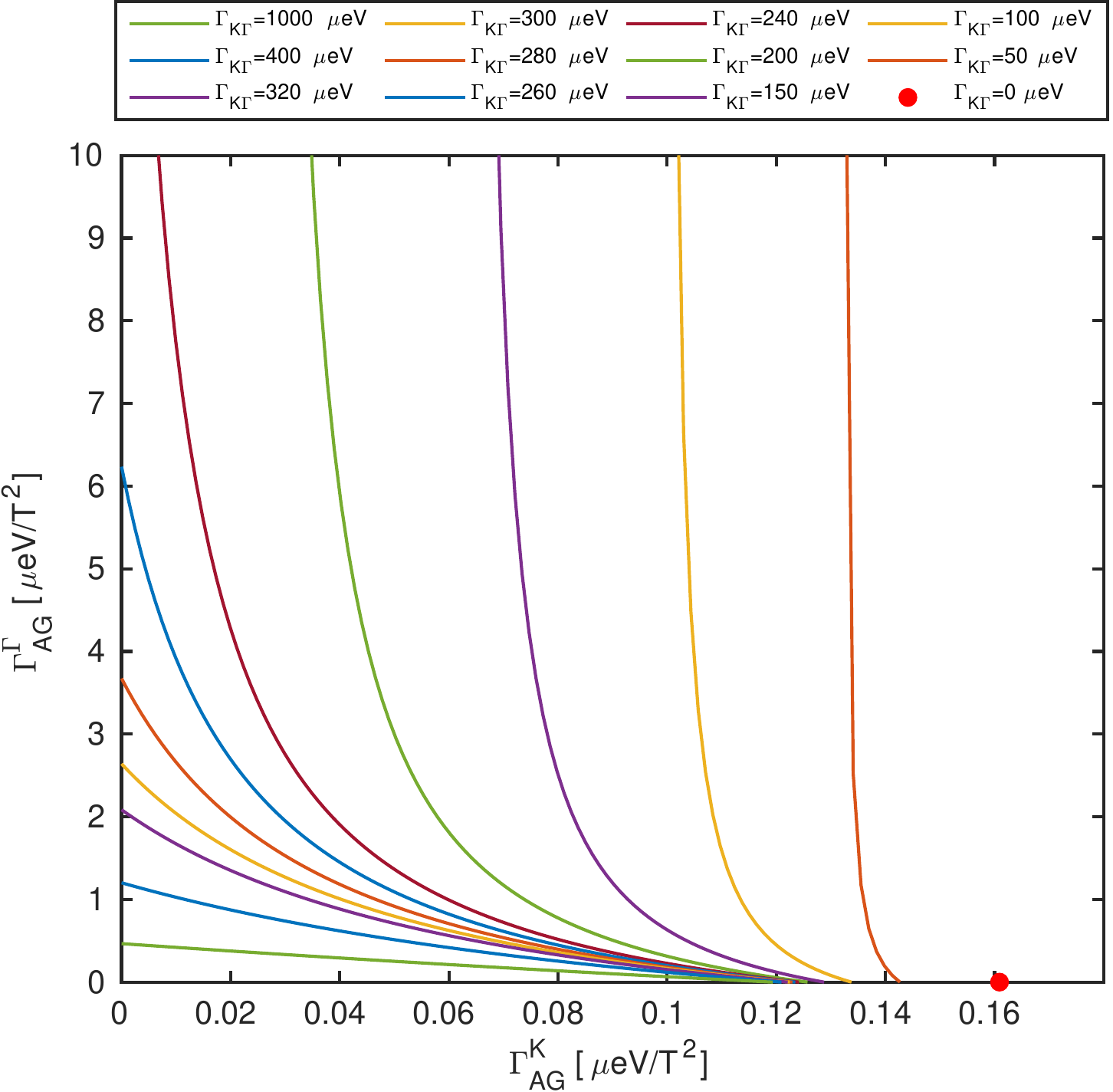}
	\caption{The values of $\Gamma_{AG}^K$ and $\Gamma_{AG}^\Gamma$ which result in $H_c=30\mathrm{T}$, for several values of $\Gamma_{K\Gamma}$ ($T=1.25\mathrm{K}$).}
	\label{fig:gammaAGHC30}
\end{figure}

At finite fields the values of the depairing parameters in both pockets play a significant role. Here they are parametrized as $\Gamma_{AG}^K(H)=\Gamma_{AG}^K H^2$ and $\Gamma_{AG}^\Gamma(H)=\Gamma_{AG}^\Gamma H^2$. Equation \eqref{eq7} can be used to determine the values of $\Gamma_{AG}^K$ and $\Gamma_{AG}^\Gamma$ such that the critical field matches the experimentally determined one $H_c=30\mathrm{T}$. No matter what the inter-pocket coupling is, a solution will always exist where $\Gamma_{AG}^\Gamma=0$ and $\Gamma_{AG}^K=\Gamma_c^{AG}(\Gamma_{K\Gamma})$, which satisfies the $H_c$ constraint. At zero coupling this is the only solution. Figure \ref{fig:gammaAGHC30} shows $\Gamma_{AG}^\Gamma$ vs $\Gamma_{AG}^K$, for different coupling strengths. Above a certain coupling strength the experimental $H_c$ can reproduced by depairing in either pocket, but below this threshold $H_c=30\mathrm{T}$ can be obtained only with a non-zero value of $\Gamma_{AG}^K$. 

Consequently, the parameter space can be reduced to just two parameters: $\Gamma_{K\Gamma}$ and $\Gamma_{AG}^K$. We can now proceed and compare the $\Delta(H)$ curves with the experimental one.

\begin{figure}[!htb]
	\centering
	\begin{minipage}{.48\textwidth}
		\centering
		\vspace*{-4mm}
		\includegraphics[width=\textwidth]{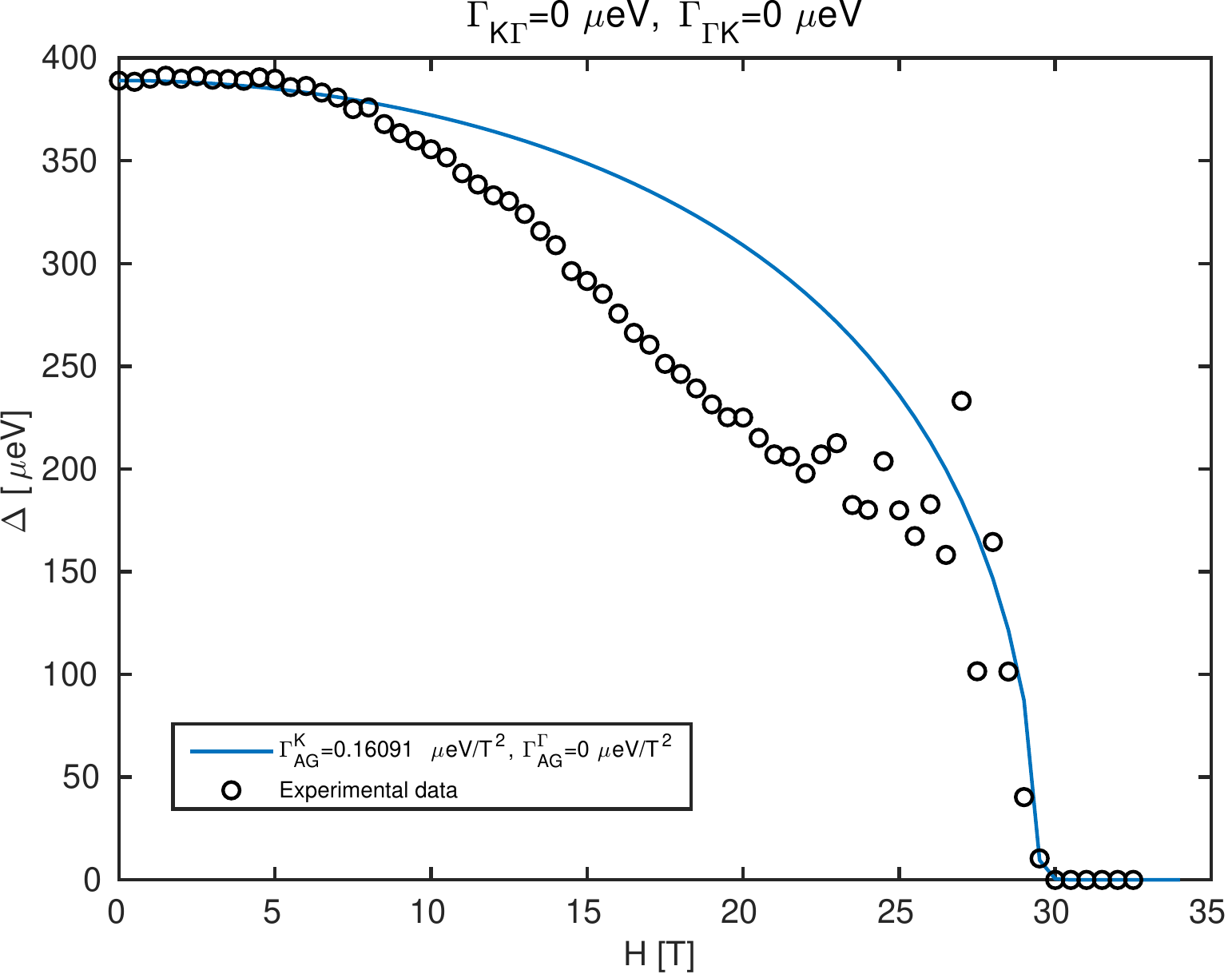}
		\caption{$\Delta$ vs $H$ curve for uncoupled pockets. The depairing is in the $K$ pocket.}
		\label{fig:deltaH_low}
	\end{minipage}%
	\begin{minipage}{0.48\textwidth}
		\centering
		\includegraphics[width=\textwidth]{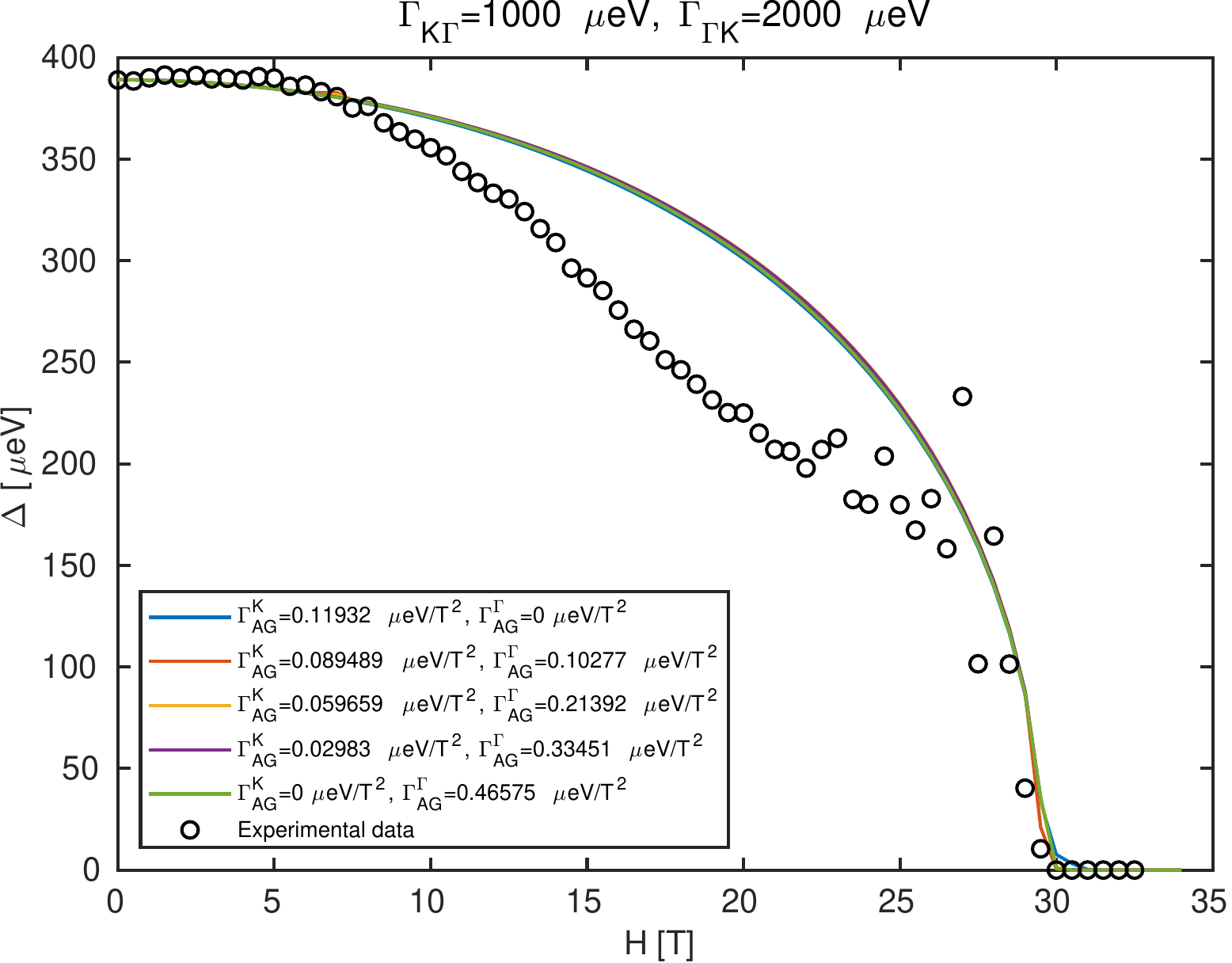}
		\caption{$\Delta$ vs $H$ curve for highly coupled pockets. The depairing is either in the $K$ pocket (blue line), the $\Gamma$ pocket (green line), or both (other lines).}
		\label{fig:deltaH_high}
	\end{minipage}
\end{figure}

In either the limit of zero coupling or very high coupling ($\Gamma_{K\Gamma}>>\Delta$) the field dependence reduces to the Abrikosov-Gorkov one. This makes sense, as at high couplings the bands are indistinguishable (i.e. DOS in both pockets is the same - the green traces in figures \ref{fig:dosK} and \ref{fig:dosG}), so depairing in either pocket can kill superconductivity, and therefore leads to the same $\Delta(H)$ trace.
Figures \ref{fig:deltaH_low} and \ref{fig:deltaH_high} show the zero coupling case (i.e. AG dependence) and the highly coupled one (for several combinations of $(\Gamma_{AG}^K,\Gamma_{AG}^\Gamma)$, which all overlap) respectively.

\begin{figure}[!htb]
	\centering
	\begin{minipage}{.48\textwidth}
		\centering
		\includegraphics[width=\textwidth]{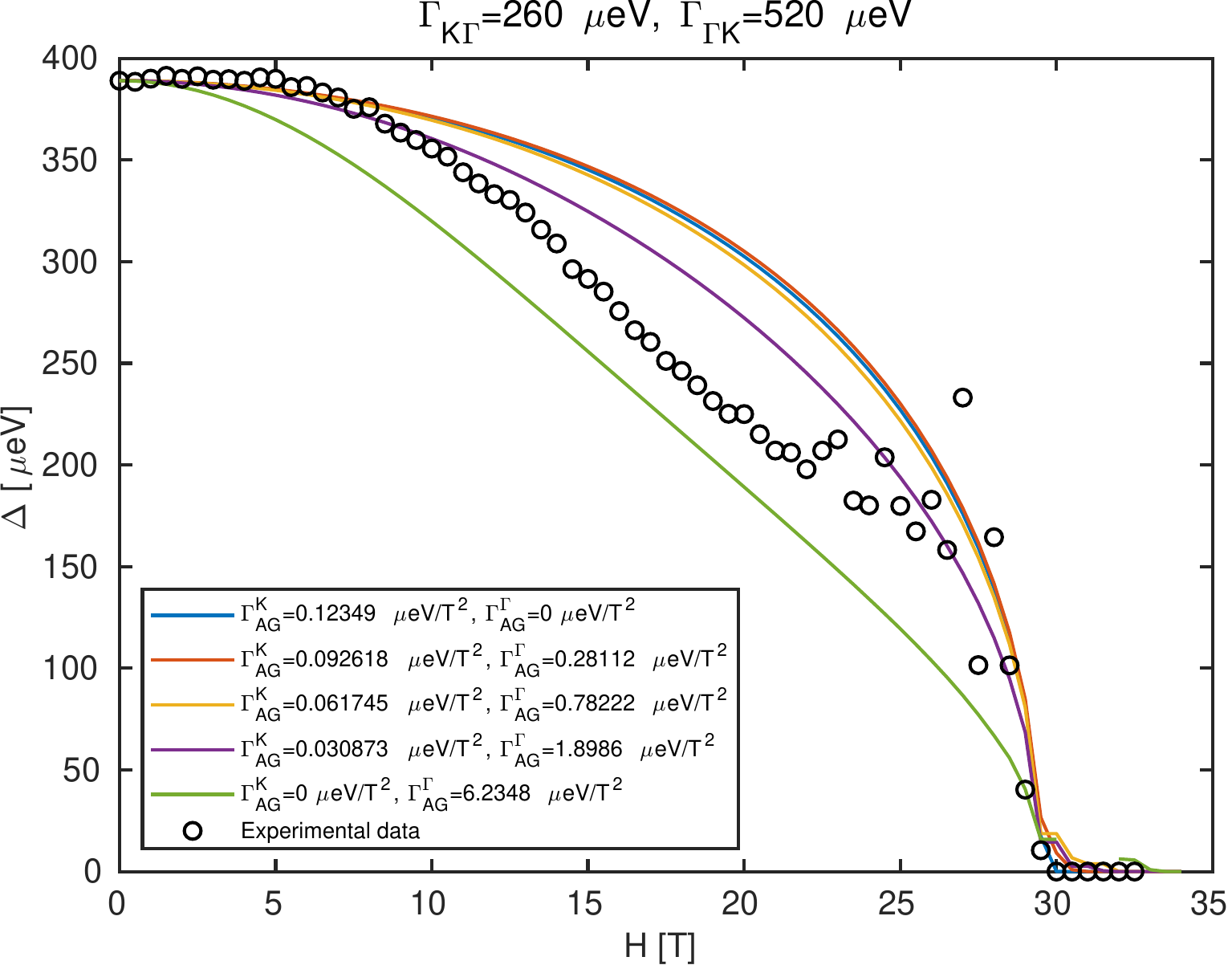}
		\caption{$\Delta$ vs $H$ curve for $\Gamma_{K\Gamma}=260\mathrm{\mu eV}$ and $\Gamma_{K\Gamma}=520\mathrm{\mu eV}$.}
		\label{fig:gamma260}
	\end{minipage}%
	\begin{minipage}{0.48\textwidth}
		\centering
		\includegraphics[width=\textwidth]{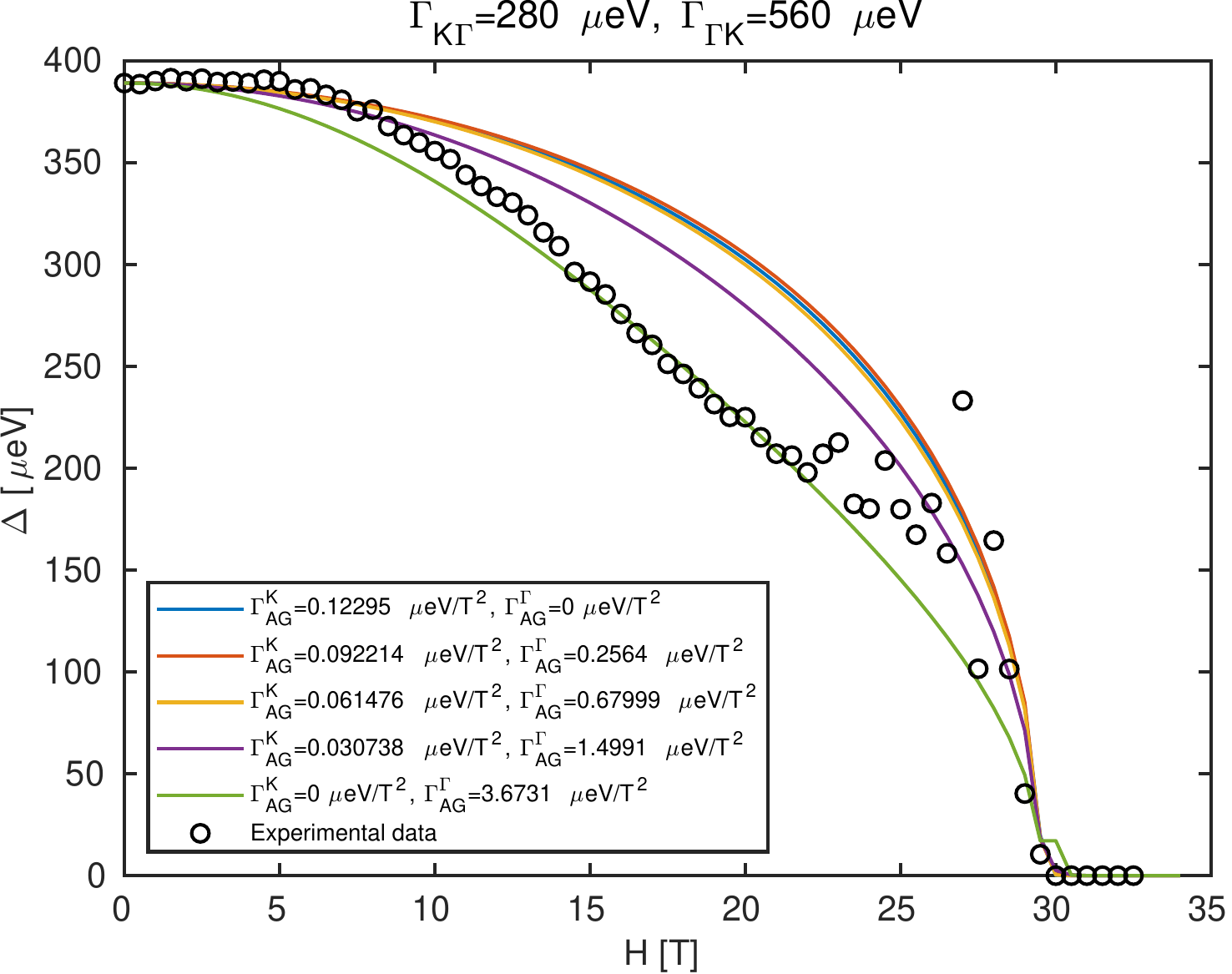}
		\caption{$\Delta$ vs $H$ curve for $\Gamma_{K\Gamma}=280\mathrm{\mu eV}$ and $\Gamma_{K\Gamma}=560\mathrm{\mu eV}$.}
		\label{fig:gamma280}
	\end{minipage}
\end{figure}

\begin{figure}[!htb]
	\centering
	\begin{minipage}{.48\textwidth}
		\centering
		\includegraphics[width=\textwidth]{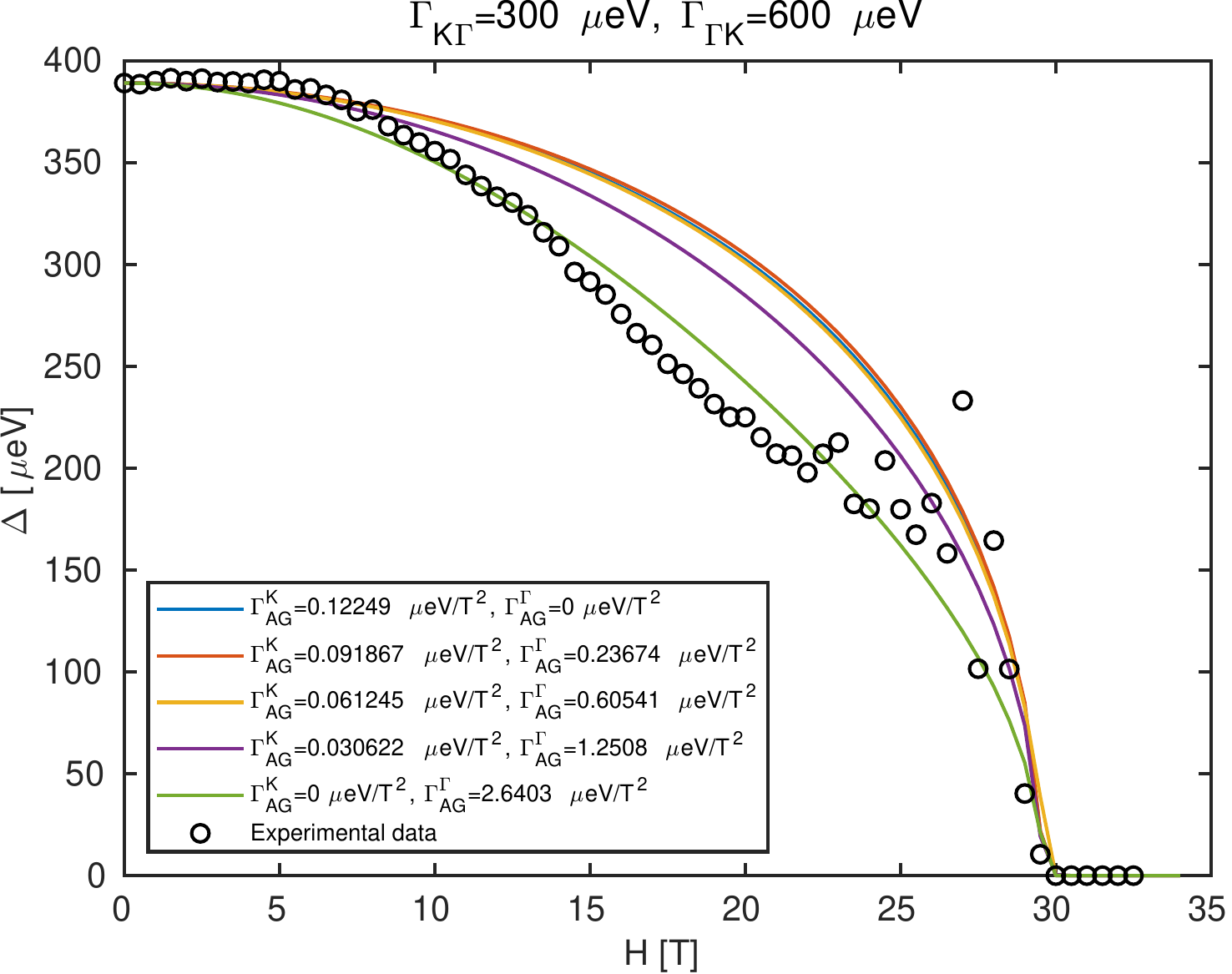}
		\caption{$\Delta$ vs $H$ curve for $\Gamma_{K\Gamma}=300\mathrm{\mu eV}$ and $\Gamma_{K\Gamma}=600\mathrm{\mu eV}$.}
		\label{fig:gamma300}
	\end{minipage}%
	\begin{minipage}{0.48\textwidth}
		\centering
		\includegraphics[width=\textwidth]{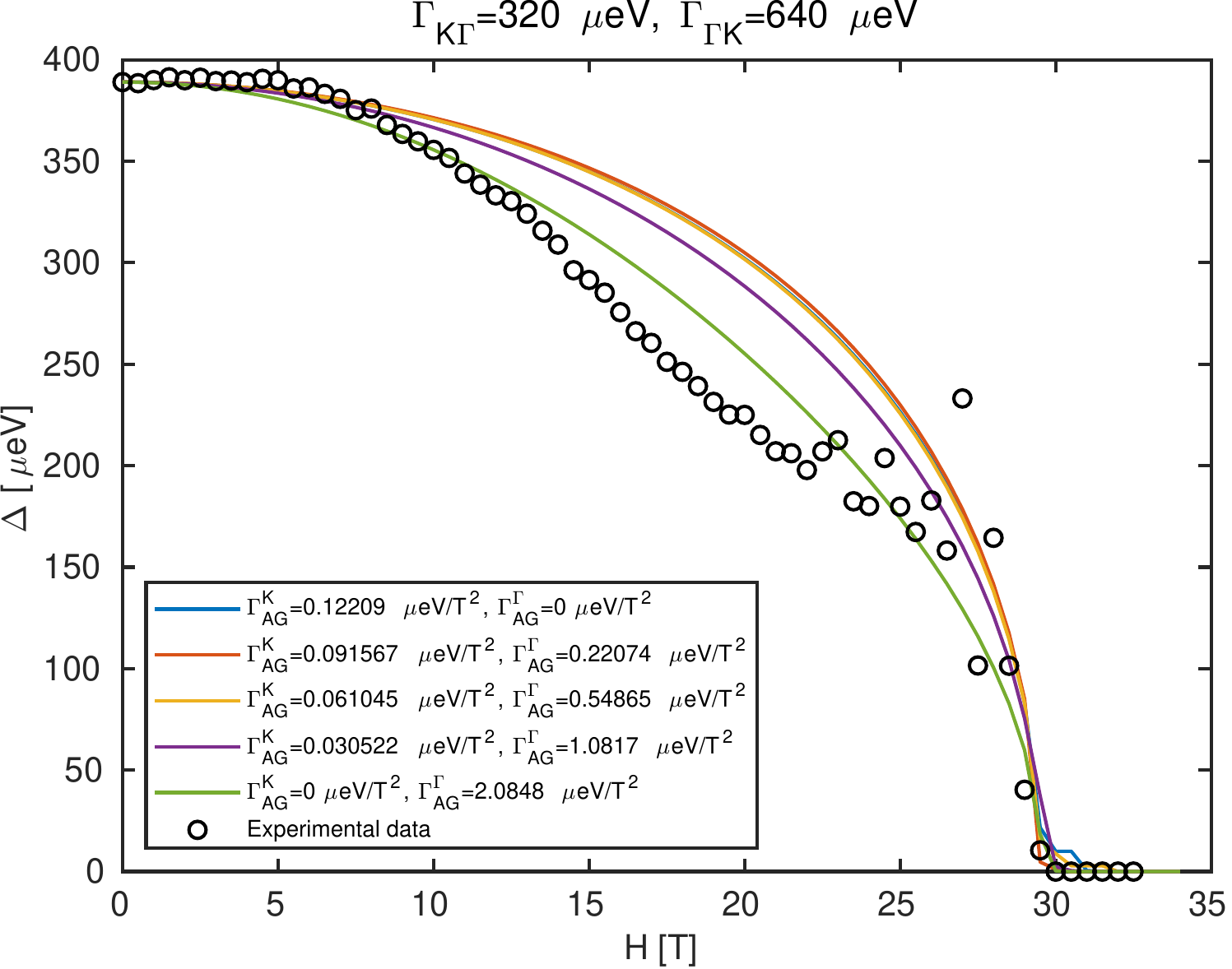}
		\caption{$\Delta$ vs $H$ curve for $\Gamma_{K\Gamma}=320\mathrm{\mu eV}$ and $\Gamma_{K\Gamma}=640\mathrm{\mu eV}$.}
		\label{fig:gamma320}
	\end{minipage}
\end{figure}

Next, the case of intermediate coupling will be discussed, for several values of $\Gamma_{K\Gamma}$,  and compared with our experimental data.\\
Figure \ref{fig:gamma260} shows $\Delta(H)$ for $\Gamma_{K\Gamma}=260\mathrm{\mu eV}$: The traces for which the depairing is dominantly in the $K$ pocket (e.g. the blue trace) agree with the data close to $H=0$ and $H=H_c$, but not in between. If the depairing is only in the $\Gamma$ pocket the trace lies below the experimental data for all fields. The purple trace shows the highest value of $\Gamma_G$ which still fits the low and high field data, increasing it further will improve the fit at intermediate fields, but at the expense of the former two.\\
Figure \ref{fig:gamma280} shows $\Delta(H)$ for $\Gamma_{K\Gamma}=280\mathrm{\mu eV}$: The green trace, for which the depairing is solely in the $\Gamma$ pocket, reproduces the experimental data relatively well, except for the bump visible at $H>20\mathrm{T}$. The "revival" of superconductivity, relative to the previous case, can be understood in the following way - the pockets are coupled strongly, superconductivity originates in the $K$ pocket while the depairing is induced in the $\Gamma$ pocket, and a smaller depairing strength is needed to obtain the same critical field. Lower depairing rates in $\Gamma$ (and higher in $K$) all result in traces above the experimental data.\\
Figure \ref{fig:gamma300} shows $\Delta(H)$ for $\Gamma_{K\Gamma}=300\mathrm{\mu eV}$: The same qualitative behavior is observed as for $\Gamma_{K\Gamma}=280\mathrm{\mu eV}$, but with a slightly worse agreement at intermediate fields.
Figure \ref{fig:gamma320} shows $\Delta(H)$ for $\Gamma_{K\Gamma}=320\mathrm{\mu eV}$: all $\Delta(H)$ curves now lie above the experimental data.

We conclude that if the two pocket model is to explain the experimental $\Delta(H)$ curve the appropriate scattering rate is close to $\Gamma_{K\Gamma}\approx280\mathrm{\mu eV}$.

\FloatBarrier

\subsubsection{Comments and discussion}\label{sec:mcmillan_comments}

Based on the analysis presented in this section, a McMillan two-pocket model can, in principle, reproduce the $\Delta(H)$ curve measured in this work, with a large inter-pocket coupling. 
For the $K\Gamma$ coupling strength required to explain the $\Delta(H)$ data the DOS of the $K$ pocket, which is dominantly probed in tunneling experiments \cite{zhu,wang2012}, is given by the black trace in figure \ref{fig:dosK}. The coherence peak reaches up to $\approx 1.4N_F$, which is comparable to our spectra measured at a finite temperature and with additional broadening, both of which reduce the height of the observed coherence peak. The shoulder, which is clearly present in the theoretical DOS, is absent from tunneling data measured with high-quality junctions - see figure 2c of the main text as well as the bi-layer tunneling spectrum from \cite{khestanova2018unusual}, reproduced here in figure \ref{fig:geim2L}.

Additionally, in thin samples we observe the BCS ratio of $\Delta/T_c\approx 1.76$ (c.f. fig. \ref{fig:GvsT}), while for the coupling needed to fit the $\Delta(H)$ curve the ratio is $\approx 2.1$ (c.f. fig \ref{fig:TcvsGamma}).

Thus, based on spectroscopic and transport data, we argue that the two pocket model with a type McMillan coupling\cite{mcmillan} is an unlikely explanation of the observed data.

\FloatBarrier

\section*{References}
\bibliographystyle{naturemag}
\bibliography{Bibliography.bib}